\documentclass[twocolumn]{aastex63}
\usepackage{amsmath}
\usepackage{upgreek}
\usepackage{CJK} 
\usepackage{color}

\newcommand{\be}{\begin{equation}}
\newcommand{\ee}{\end{equation}}
\def\lta{\,\raise 0.3 ex\hbox{$ < $}\kern -0.75 em
 \lower 0.7 ex\hbox{$\sim$}\,}
\def\gta{\,\raise 0.3 ex\hbox{$ > $}\kern -0.75 em
 \lower 0.7 ex\hbox{$\sim$}\,}

\received{March 22, 2021}
\revised{April 24, 2021}
\accepted{May 11, 2021}
\submitjournal{ApJ}
 
\begin{document}
\begin{CJK*}{UTF8}{gbsn} 

\title{Orbital migration and circularization of tidal debris by Alfv{\'e}n-wave drag: \\ circumstellar debris and pollution around white dwarfs}

\correspondingauthor{Shang-Fei Liu, co-first author}
\email{liushangfei@mail.sysu.edu.cn; shangfei.liu@gmail.com}

\author[0000-0003-4045-9046]{Yun Zhang (张韵)}
\affil{Universit\'e C\^ote d'Azur, Observatoire de la C\^ote d'Azur, CNRS, Laboratoire Lagrange, Nice 06304, France}

\author[0000-0002-9442-137X]{Shang-Fei Liu (刘尚飞)}
\affil{School of Physics and Astronomy, Sun Yat-sen University, 2 Daxue Road, Tangjia, Zhuhai 519082, 
Guangdong Province, China}
 
\author[0000-0001-5466-4628]{Douglas N.C. Lin (林潮)}
\affiliation{Department of Astronomy and Astrophysics, University of California, Santa Cruz, USA}
\affiliation{Institute for Advanced Studies, Tsinghua University, Beijing, 100086, China}

\begin{abstract}
    A significant fraction of white dwarfs (WDs) exhibit signs of ongoing accretion of refractory elements at rates $\sim10^3$--$10^7$ kg s$^{-1}$, among which, 37 WDs were detected to harbor dusty debris disks. Such a concurrence requires not only fertile reservoirs of planetary material, but also a high duty cycle of metal delivery. It has been commonly suggested that this material could be supplied by Solar System analogs of Main Belt asteroids or Kuiper Belt objects. Here we consider the primary progenitors of WD pollutants as a population of residual high-eccentricity planetesimals, de-volatilized during the stellar giant phases. Equivalent to the Solar System's long-period comets, they are scattered to the proximity of WDs by perturbations from remaining planets, Galactic tides, passing molecular clouds, and nearby stars. These objects undergo downsizing when they venture within the tidal disruption limit. We show quantitatively how the breakup condition and fragment sizes are determined by material strength and gravity. 
    Thereafter, the fragments' semi-major axes need to decay by at least $\sim$6 orders of magnitude before their constituents are eventually accreted onto the surface of WDs. 
    We investigate the orbital evolution of these fragments around WDs 
    and show that WDs' magnetic fields induce an Alfv{\'e}n-wave drag during their periastron passages and rapidly circularize their orbits. This process could be responsible for the observed accretion rates of heavy-elements and the generation of circum-WD debris disks. A speculative implication is that giant planets may be common around WDs' progenitors and they may still be bound to some WDs today. 
    
\end{abstract}

\keywords{White dwarf stars (1799) --- Minor planets (1065) --- Exoplanet evolution (491) --- Exoplanet dynamics (490)--- Debris disks (363)}
 
\section{Introduction}
\label{sec:intro} 

    After losing their red giant progenitors' envelope, a significant fraction of white dwarfs (WDs) were detected to harbor residual planetary bodies, as revealed by the evidence of metal pollution in their spectra \citep{Zuckerman2003, Koester2014}, infrared excesses and metal emission lines from close-in debris disks \citep{Zuckerman1987, Barber2012, Manser2020}, and photometric and spectroscopic signatures of transiting planetary material in close proximity to the WDs \citep{Vanderburg2015, Xu2017, Manser2019, Vanderbosch2020}.  These phenomena can be readily explained by the tidal disruption and accretion of remnant planetary bodies (such as asteroids and comets) that were perturbed into nearly parabolic orbits with pericenters in proximity to their host WDs \citep{Debes2002, Jura2003, Kratter2012, Veras2015a, Chen2019, Malamud2020}.   However, detailed dynamical processes connecting these planetary remnants around WDs with the planetary systems in the late stages of stellar evolution remain poorly understood \citep[e.g.,][and references therein]{Veras2016a}.  
    
    Due to the mass loss and size expansion of a host star in its post-main-sequence evolution, the orbiting objects migrate outward \citep{Hadjidemetriou1963, Adams2013} and some may be engulfed by the enlarging star on the giant branch\citep{Schroder2008}. These stars' intensified luminosity also heats the surface of residual objects, sublimates and disperse their volatile contents. With the consideration of the time-varying giant-branch luminosities and the consequently intensified Yarkovsky effect, \citet{Veras2019a} found 
    the engulfment survivors generally have asymptotic orbital semi-major axes larger than $10$--$100$ au.  In order to form the observed circumstellar debris disks in the proximity of the WDs (with orbital semi-major axes on the order of one solar radius and small orbital eccentricities), substantial orbital migration and circularization of these objects and their tidal fragments are required.
    
    A commonly held view is that remnant planetary bodies are grounded to $\upmu$m-sized dust by WDs' tidal effect \citep{Jura2003} and 
    by collisional cascades \citep{Kenyon2017} when they approach the WDs' tidal radii.  Subsequently, the $\upmu$m-sized and smaller dust can be circularized near the tidal radius and be accreted onto the WDs by their radiation \citep{Rafikov2011, Veras2015a} and/or magnetic drag \citep{Hogg2021} on relatively short timescales.

        \begin{figure*}[htbp]
        \centering
        \includegraphics[width = 18 cm]{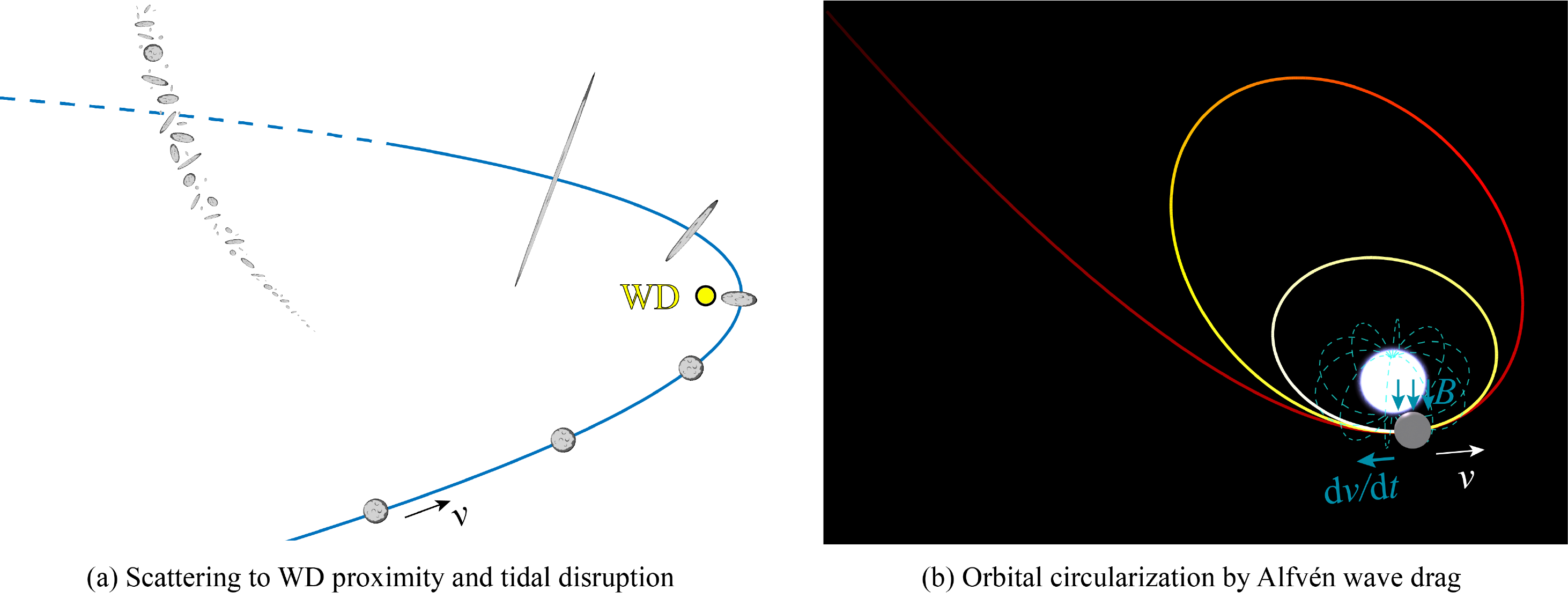}
        \caption{A schematic of the dynamical evolution of remnant planetary bodies around a magnetic WD.}
        \label{f:schematic}
        \end{figure*}

    However, there is hardly any empirical constraint on the particle size distributions in the circumstellar debris of WDs \citep{Farihi2016} and it is likely that cm-sized grains to even km-sized objects could be common \citep{Rafikov2012, Vanderburg2015, Xu2017, Manser2019}.  In effect, the size distribution of tidal disruption debris highly depends on the granular constituents and material strength of the progenitor body.  Wide-ranging observations and \textit{in-situ} explorations have confirmed that most of Solar System small bodies, with diameters $\gtrsim$200 m and $\lesssim$50 km (which belong to one of the most common small body population in the Solar System), have rubble-pile structures, in which boulders and grains are held together mainly by gravitational forces \citep{Richardson2002, Hestroffer2019}. As revealed by recent space missions, the size of boulders and grains of rubble-pile bodies typically ranges from sub-cm to 100 m \citep[e.g.,][]{Walsh2019} with individual material tensile strength on the order of $0.1$--$10$ MPa \citep{Pohl2020}.  This level of material strength can help some large boulders to remain intact within the strength-free Roche radius \citep{Holsapple08, ZhangLin2020}.  Therefore, the tidal debris of a rubble-pile body at WDs may have size distributions close to the size distribution of its constituents.  
    
    For a 100-km-sized body with an internal viscosity similar to that of molten rocks, tidal friction can circularize its orbit within appropriate timescales \citep{OConnor2020}.  However, the stiff interior of the cm-sized grains to 100-m-sized boulders can not provide the required internal viscosity for tidal friction migration. WD radiation \citep{Rafikov2011, Veras2015a} and associated material sublimation \citep{Veras2015b} are inefficient to circularize the orbits of decimeter-sized or larger particles.  Therefore, other mechanisms are needed to account for the high-eccentricity migration of objects within this size range.
    
    Around 8--20\% of WDs currently have magnetic fields with strengths ranging from $10^3$ to $10^9$ G \citep{ferrario2005,  Ferrario2015}.  The origin of WDs' magnetic fields remains an outstanding issue \citep{Isern17}.  Observations \citep[e.g.,][]{Hollands2015, Kawka2019} suggest a higher incidence of magnetism in WDs at low effective temperatures, which may indicate that magnetic fields has been amplified during the evolution of WDs.  The presence of magnetic fields has a significant influence on the dynamics of $\upmu$m-sized dust around WDs \citep{Farihi2017, Hogg2021} and the consequent Ohmic heating and Lorentz drift can cause the decay of close-in orbits of asteroid-sized objects \citep{Bromley2019, Veras2019b}.  In addition, the Alfv{\'e}n wave emitted from an object orbiting a magnetic WD might also cause significant drag effects on its orbit \citep{Drell1965}, which will be examined in this study.
    
    Here we investigate the role of material strength on the tidal fragmentation size distribution and examine the orbital migration of tidal debris by the Alfv{\'e}n-wave drag in the proximity of WDs.  The illustrative Figure \ref{f:schematic} presents the two main processes considered in this study that attribute to the formation of circumstellar debris and pollution of WDs.  We briefly recapitulate the dynamical evolution of fragments' parent bodies under this scenario in Section \ref{sec:lpcs}. We establish, in
    Section \ref{sec:materialstrength}, the tidal disruption condition around WDs and derive the survivor size for various material strength and pericenter distances.  In 
    Section \ref{sec:alfven}, we analyze the orbital energy dissipation and migration of survivors within the possible size range under the effect of Alfv{\'e}n-wave drag. 
    In Section \ref{sec:numerical}, dedicated numerical simulations are carried out to validate our analytic approximation.  
    Taking these consideration together, we elaborate, in Section \ref{sec:lpctodebris}, 
    the scenario of the transformation of long-period comets (LPCs) into circular debris close to the WDs.
    We summarize our results and discuss their implications regarding circumstellar debris and pollution at WDs 
    in Section \ref{sec:summary}.

\section{High-eccentricity migration of heavy-elements' parent bodies}
\label{sec:lpcs}

In this paper, we adopt the scenario, as shown in Fig.~\ref{f:schematic}, that debris disks and accreted refractory elements at WDs originated from remnant planetary bodies residing in regions analogous to the Kuiper belt and Oort cloud in the Solar System
\citep{veras2011, Veras2020a}. We first recapitulate several bottlenecks associated with this scenario: 
1) an adequate reservoir of residual planetesimals, 2) the conversion from distant residual planetesimals to 
LPCs, 3) a sustained supply of mostly-refractory cometary bodies over Gyr time scales, and 4) the orbital decay of 
the cometary nuclei, and their fragmentation and accretion onto the WDs.  We briefly discuss some of these issues 
in the present context.  

\subsection{Residual planetesimal belts around WDs.}
\label{sec:reservoir}
Since the observed properties of comets in the Solar System can be characterized in detail, 
their origin and evolution \citep{oort1950} are often used as an analogue model and guide for the reservoirs 
of residual planetesimals around other stars including WDs and their main-sequence
progenitors.  A frequently applied minimum-mass solar-nebula (MMSN) model for the formation 
of the Solar System is based on the assumptions that major, dwarf, and minor planets are 
formed {\it in situ} with a maximum retention efficiency of heavy elements \citep{Hayashi85}.  
Its extrapolation implies a mass $\sim 10^2 M_\oplus$ of planetesimals building 
blocks within a few tens of au in the solar nebula.  Protoplanetary and debris disks with similar 
inventory and dimension appear to be prevalent among young stars 
\citep[e.g.,][and references therein]{andrews2018, long2018}. The progenitors of currently observed WDs are generally more massive than the Sun with a mass distribution peaking at 1.5--2.5$M_\odot$
\citep{hurley2000, Tremblay2016, Barrientos2021}. 
Since 1) exoplanets, especially multiple super-Earth systems are ubiquitous around FGK main-sequence stars and 2) the 
occurrence rate of gas giant planets increases with their host star's mass \citep{cumming2008, howard2012, 
petigura2013, fabrycky2014}, we assume that planet formation processes are robust \citep{ida2004, ida2013} 
and the initial seeding as well as subsequent dynamical evolution of residual planetesimals in the Solar 
System can be applied to planets and their building blocks around WDs and their progenitors.

During the advanced stages of their  growth, the emerging proto ice and gas giant planets in the 
Solar System scattered nearby planetesimals into the Kuiper belt (with semi-major axes, 30 au $ \lta a \lta 10^3$ au) and inner
($a \sim 10^{3-4}$ au) and outer ($a \sim 10^{4-5}$ au) Oort cloud, respectively \citep{oort1950, 
Zhoulin07, dones2015}. As their natal disks deplete, eccentric planets' 
secular resonances sweep through the planet forming region, exciting the eccentricities of residual planetesimals, efficiently clearing them from the planetary domain (including the asteroids' main belt),  
and launching them to much larger (or smaller) semi-major axes \citep{nagasawa2005, zheng2017a, zheng2017}. 
These planetesimal scattering processes also drive the planets to migrate
\citep{fernandez1984}. After the disk is severely depleted, 
the dynamical evolution of residual planetesimals continues to be influenced by the secular evolution 
and stability of their hosting planetary system \citep{tsiganis2005, gomes2005}.   

Since most stars, including the Sun, are formed in dense stellar clusters that become
dispersed (under the combined contribution from the stellar relaxation and Galactic tidal field) 
within $\sim$0.1 Gyr \citep{adams2010}, they withstand close encounters with stellar neighbors, which also lead to 
modest semi-major axis changes and a nearly uniform eccentricity and inclination distribution for 
the planetesimals in the inner Oort cloud 
\citep{Veras2020a}. A significant fraction of more
distant residual planetesimals become detached from their host stars \citep{Portegies18}.

During the Sun's main-sequence evolution, it has been repeatedly bombarded by a few 
short-period (SPCs with orbital periods $ \leq 200$ yr, e.g., Halley) and long-period (LPCs with orbital periods $ \geq 200$ yr, e.g., 
Hale-Bopp) comets each year.  The SPCs originate from the Kuiper belt which stores a total of 
$\lesssim 10^{-3} M_\oplus$ dwarf planets and residual planetesimals \citep{fraser2014} and it is being 
continually sculpted by perturbation from the gas and ice giants \citep{duncan1987, duncan1988}.  
The LPCs are drawn from the scattered disk and Oort cloud. Under the perturbation of Galactic 
tides and passing giant molecular clouds (GMCs) and stars, residual planetesimals in this region generally preserve 
their large semi-major axes \citep{spurzem2009, punzo2014, Veras2020a} while their eccentricities are excited 
over the host stars' multi-Gyr main-sequence evolution (see Section \ref{sec:eccexc}).

The Sun is half way through its 10 Gyr main-sequence evolution.  
During a brief interval ($\sim 1$ Myr) on the asymptotic-giant branch and planetary-nebula phase, 
the progenitors of WDs out-shined the Sun by 3--5 orders and lost a major fraction of their original 
mass \citep{hurley2000}.  Modest size (up to 10 km in radius) asteroids originally located in the current main belt 
region ($\sim 2-4$ au) are either engulfed by the greatly expanded stellar envelope    
or downsized through sublimation.  Kuiper belt objects (KBOs, initially located at $30$ au to a few $10^2$ au) 
are de-volatilized and their orbital semi-major axes undergo adiabatic expansion. Sub-km KBOs also migrate due to 
the Yarkovsky and the coupled YORP effects \citep{Veras2019a}.  Neither the composition nor the orbits of 
residual planetesimals in the outer Kuiper belt and inner Oort cloud (originally at or beyond a few $ 10^2$ au) 
are altered largely by host stars' intensified radiative flux.  In response to the stellar mass loss during the
subsequent planetary nebula phase, most planets and residual planetesimals with initial 
$a \lesssim 10^{3}$ au are retained by doubling their semi-major axes  
without changing their eccentricity \citep{Debes2002, Adams2013}.  However, more distant planetesimals 
with initial $a \gtrsim 10^3$ au are either marginally retained with high 
eccentricities or released from the host stars' gravitational confine \citep{veras2011, Rafikov18}.  Depending 
on the progenitors' initial mass and planetesimals' eccentricity and distribution, the region beyond $\sim 10^4$ au 
in recently emerged (young WDs may be sparsely populated with a precipitous drop-off in their density distribution.
 
The residual planetesimals around mature WDs continue to be perturbed by accompanied 
planets, Galactic tides, and passing GMCs and stars. But, their overall kinematic distribution evolves 
slowly on time scales longer than the Gyr age of their WD hosts. With these considerations, 
we assume that there is a population of residual planetesimals with asymptotic semi-major axes $a \lesssim  10^{4}$ au 
(equivalent to the region in the Solar System's Kuiper belt and Oort cloud) around WDs.  In order 
for these objects to reach the tidal radius of their WD hosts, they must acquire nearly parabolic orbits in which eccentricity 
$e = 1- \varepsilon$ with $\varepsilon \lesssim10^{-6}$.

\subsection{Excitation of nearly parabolic orbits}
\label{sec:eccexc}
In the outskirts of the Solar System, secular torque from Galactic tides and passing GMCs and stars
leads to changes in these residual planetesimals' angular momentum, excitation of their eccentricity 
$\Delta e$, and randomization of their inclination with much smaller fractional changes in their 
energy $\Delta E/E$ and semi-major axes $\Delta a/a$ \citep{heisler1986}. Consequently, their pericenter 
distances $q$ diffuse towards the surfaces of their host stars with a nearly uniform inclination distribution 
\citep{wiegert1999, boe2019}.  
Due to the host stars' gravity, residual planetesimals with modest $a$ (comparable to that in the Kuiper 
belt and inner Oort cloud) suffer limited fractional changes 
in their eccentricities $\Delta e/e$ and pericenter distances $\Delta q/q$ per orbit. A flux of residual planetesimals is continually fed to
the proximity of planet-free host stars as LPCs. Since the strength of external perturbation relative
to host stars' gravity is an increasing function of $a$, many of the injected LPCs 
originated from the outer region where most of the residual planetesimals are 
populated (with $a \sim$ a few $10^3$ au).  

However, rich populations of residual planetesimals at Oort cloud distances is synonymous with not only 
ample supplies of planet-building blocks but also the omnipresence of giant-planet scatterers (Section \ref{sec:reservoir}). 
Around stars with giant planets at locations similar to those in the Solar System, the contraction of planetesimals' 
$q$ is temporarily interrupted by planets' perturbation just exterior to their orbits \citep{fernandez1981}.  
The giant planets' gravity also imparts changes in the energy and angular momentum to the planetesimals 
and increases in their $a$ \citep{levison2006}.  With sufficiently large apo-center distances 
($\gtrsim 10^4$ au) where the stellar gravity is relatively weak, Galactic tides and passing 
stars as well as GMCs induce the residual planetesimals to undergo large incremental $\Delta q/q$. These persistent 
changes allow a fraction of them to cross the giant planets' 
orbital barrier and be transformed into Sun-grazing LPCs \citep{hills1981, Kaib2009}.

During the Sun's lifespan, these competing effects cause about half of residual planetesimals population to be lost to 
the interstellar space and a smaller fraction to be injected into the inner Solar System \citep{Hanse18}.  With a 
total estimated population $N_{\rm rp} \sim 10^{12}$ of residual planetesimals in the outer Oort cloud, this 
scenario can account for the observed infusion rate $\dot{n}_{\rm LPC} \sim$ a few events per year \citep{Everhart67, 
Francis05, Neslusan07, Kaib2009}. The inner Oort cloud replenishes the outer Oort cloud with residual 
planetesimals\citep{hills1981}. Although a much larger population of residual planetesimals has been hypothesized for the
inner Oort cloud \citep{fernandez1997, levison2001}, it does not directly source LPCs such that there has not been any 
observational constraints
on its inventory. From the average mass ($M_{\rm LPC} \sim 4 \times 10^{13}$ kg) and size distribution \citep{Weissman97, 
meech2004} of individual LPCs, the corresponding mass flux is ${\dot M}_{\rm LPC} \sim M_{\rm LPC} \dot{n}_{\rm LPC} \sim 
10^{7}$ kg s$^{-1}$ and the equivalent total mass of this residual planetesimals in the outer Oort 
cloud is $M_{\rm rp}= N_{\rm rp} M_{\rm LPC} \sim 2-7 M_\oplus$ \citep{weissman1983, Weissman97}. Uncertainties in these mass estimates are of order unity. Recent PanSTARR survey confirms the extrapolated
existence of a rich population of residual planetesimals with uniform inclination distribution, random orbital 
orientations, and comparable total mass in the outer Oort cloud \citep{boe2019}.

Around other planet-bearing stars, including WDs, the transition radii between SPCs and LPCs and between
inner and outer Oort clouds are smaller for more compact or less massive planetary systems or in more
dense stellar environments \citep{fernandez1997}. On Gyr time scales, residual planetesimals with $a \sim 10^{3-4}$ au
can continually feed their host stars at rates similar to $\dot{n}_{\rm LPC}$ in the Solar System whether 
or not they have accompanying giant planets. The magnitude of ${\dot M}_{\rm LPC}$ is comparable to or 
slightly larger than those inferred for the WDs' accretion (${\dot M}_{\rm WD} \sim 10^5-10^6$ kg s$^{-1}$) of refractory 
elements \citep{Manser2019, Chen2019}. In contrast to those around the Sun, the LPCs around WDs may have 
lost their volatile content 
during the red giant phase.  Over the multi-Gyr age of typical WDs, the integrated amount of mostly refractory 
elements accreted onto them, $M_{\rm Z} \sim 0.01-0.1 M_\oplus$, is a small fraction of the present-day total mass estimated for the LPCs \citep{boe2019} and residual planetesimals ($M_{\rm rp}$) in the Oort cloud but larger than the total mass in asteroids' 
main belt and KBOs in the Solar System.  In terms of the mass budget, the outlying LPCs appear to be the most likely 
candidates for the parent bodies of WDs' dusty disks and heavy elemental contaminants.
   
LPCs in the Solar System have an average size of a few km with a steep fall off in their
size frequency distribution \citep{meech2004, boe2019}.  {\it In situ} formation of larger planetesimals  
beyond the Kuiper belt is likely to limit the high end of their mass function due to the density 
fall off of their building block material \citep{morbidelli2021}.  While the large (100--1000 km-size) 
dwarf planets may dominate the total mass of the scattered population in the Kuiper belt 
\citep{fraser2014}, the probability and resultant distribution of their asymptotic 
apo-center distances decline with their semi-major axes expansion factor.  We expect that
the number density of large residual planetesimals at $10^{3-4}$ au from their host stars 
may be so sparse that their rare transition into LPCs is negligible.  

Based on the extrapolation of these observed properties for the Solar System LPCs to the parent 
bodies of debris disks around and heavy elements accreted onto the WDs, we assume that, they 
have an average size $r\sim 10$ km, $a \sim 10^{3-4}$ au, $q \sim 0.1 R_\odot - R_\odot$, random 
distributions in inclination, pericenter argument, and ascending node.  We choose this range of $a$ for LPCs on the bases:
1) a significant fraction of the residual planetesimals at this location is retained despite
the mass loss during the red giant phase; 2) the external perturbation is sufficiently
intense for the residual planetesimals to diffuse towards the stellar surface over Gyr
time scales; 3) although giant planets at a few au, if present, would temporarily interrupt the
decline in $q$, their barrier can be effectively bypassed through dynamical channels similar to
those in the Solar System \citep[i.e.,][]{Kaib2009}.  The following sections examine the tidal downsizing and orbital evolution of these parent bodies and their fragments.

\section{Tidal downsizing of residual planetary bodies}
\label{sec:materialstrength}

There are 
two potential reservoirs for the parent bodies of the debris disks and WD pollutants. These
parent bodies may originate from the residual planetesimals either at a location interior to some retained 
giant planets (similar to the asteroids) or far from their host stars (analogous to the long-period comets, LPCs, see
Section \ref{sec:lpcs}).
Both the asteroid and the LPC scenarios assume the accreted heavy elements are mostly byproducts of collisions, PR drag, orbital decay, planetary scattering, and tidal disruption \citep[e.g.,][]{Chen2019, Veras2014, Malamud2021}. The main differences between the dynamical
properties of the parent bodies in these two models are: 1) the semi-major axis, 2) inclination, and 3) eccentricity
distribution. Here, we adopt a general model for the fragmentation of parent bodies with diverse structural properties and material strengths.

    \subsection{Condition for tidal fragmentation}
    
        As perturbed by the combination of Galactic tides, passing stars, residual giant planets, and stellar components, a fraction of residual asteroids and comets are continually scattered into nearly parabolic orbits with pericenters that are in close proximity to their host WD \citep{Kaib2009, Kratter2012, Veras2016a, Hamers2016}.  They undergo tidal disruption when the pericenter distances are reduced below some critical values, where tidal forces imposed by the WD overwhelm the gravitational forces and material strength that hold these objects together and may disrupt them.  
        
        The tidal disruption limiting distance of an initially spherical, non-spinning, viscous fluid object is \citep{Sridhar1992}
        \begin{equation}
            d_\mathrm{gra}=1.05\Big(\dfrac{M_\star}{\rho}\Big)^{1/3}, 
        \end{equation}
        where $\rho$ is the bulk density of the object and $M_\star$ is the mass of the WD.  $d_\mathrm{gra}$ is appropriate to predict the tidal disruption distance for objects in the gravity-dominated regime, such as gas giants, super-Earths with molten interiors, and minor and dwarf planets with sizes $\gtrsim 10$ km.  While for small objects in the strength-dominated regime, their intrinsic material shear and tensile strengths can prevent tidal disruption within this limit and determine their survivability \citep{ZhangLin2020, ZhangMichel2020}.  
        
        To reflect the material characteristics, we follow the pioneer work of \cite{Holsapple08} and use the  elastic-plastic continuum theory  for solid and granular material to estimate the tidal disruption limit of small bodies. For geological materials, the tensile and shear strengths depend strongly on the loading conditions \citep[i.e., circumambient pressures;][]{Holsapple2007}. The Drucker-Prager failure criterion that is appropriate for a large class of solid and granular geological material is applied to determine the failure of such objects,
        \begin{equation}
            \label{eq:dp}
            \sqrt{J_2}\le k-sI_1, 
        \end{equation}
        where $I_1$ is the first invariant of the Cauchy stress tensor, and $J_2$ is the second invariant of the deviatoric stress tensor, which can be written in terms of the principal stresses, ($\sigma_1$, $\sigma_2$, $\sigma_3$), as $I_1=\sigma_1+\sigma_2+\sigma_3$, $J_2=[(\sigma_1-\sigma_2)^2+(\sigma_2-\sigma_3)^2+(\sigma_3-\sigma_1)^2]/6$.  The material constants $s$ and $k$ can be expressed as a function of the commonly-used friction angle, $\phi$, and cohesive strength $C$ (which represents the shear strength at zero pressure), 
        \begin{equation}
          s = \dfrac{2\sin{\phi}}{\sqrt{3}(3-\sin{\phi})},~~k=\dfrac{6C\cos{\phi}}{\sqrt{3}(3-\sin{\phi})}.
        \end{equation}
        The structure of an object fails when its internal stress state violates this criterion.

        Subject to the self gravity and the host WD's tidal effect, the volume-averaged normal stress components of a non-spinning spherical object for an encounter at a distance $d$ with respect to the WD's mass center are given as
        \begin{equation}
        \begin{split}
            &\bar{\sigma}_x=\bigg[-\dfrac{4}{3}\pi\rho^2G+\dfrac{8\pi}{3}G\rho\rho_\star\Big(\dfrac{d}{R_\star}\Big)^{-3}\bigg]\dfrac{r^2}{5} , \\
            &\bar{\sigma}_y=\bigg[-\dfrac{4}{3}\pi\rho^2G-\dfrac{4\pi}{3}G\rho\rho_\star\Big(\dfrac{d}{R_\star}\Big)^{-3}\bigg]\dfrac{r^2}{5} , \\
            &\bar{\sigma}_z=\bigg[-\dfrac{4}{3}\pi\rho^2G-\dfrac{4\pi}{3}G\rho\rho_\star\Big(\dfrac{d}{R_\star}\Big)^{-3}\bigg]\dfrac{r^2}{5} , 
        \end{split}
        \end{equation}
        where $\rho$ and $r$ are the bulk density and radius of the object, and $\rho_\star$ and $R_\star$ are the bulk density and radius of the WD, respectively. $G$ is the gravitational constant. These stress components are expressed in a right-hand Cartesian coordinate system originating at the object's mass center, where $x$-axis is towards the WD center and $z$-axis is perpendicular to the orbital plane. Since the averaged shear stress components are exactly zero in this coordinate system, we have the averaged principle stresses $\bar{\sigma}_1 = \bar{\sigma}_x$, $\bar{\sigma}_2 = \bar{\sigma}_y$, $\bar{\sigma}_3 = \bar{\sigma}_z$.
        
        By substituting the stresses into the failure criterion (Eq.~(\ref{eq:dp})), the tidal disruption limit for an object in the strength-dominated regime is given as
        \begin{equation}
            \label{eq:tidallimit}
            d_\mathrm{str} = \Big(\dfrac{\sqrt{3}}{4\pi}\Big)^{1/3}\Big(\dfrac{5k}{4\pi Gr^2\rho^2}+s\Big)^{-1/3} \Big(\dfrac{M_\star}{\rho}\Big)^{1/3},
        \end{equation}
        which generally decreases with either a larger friction angle $\phi$ or stronger cohesion $C$.  This limit is much smaller than the limit in the gravity-dominated regime, e.g., even for low-strength material with $\phi=25^\circ$ \footnote{The friction angle of geological materials commonly ranges from 25$^\circ$ to 50$^\circ$ \citep[e.g.,][]{Bareither2008}.} and $C = 0$ Pa, $d_\mathrm{str} = 0.9 (M_s/\rho)^{1/3} < d_\mathrm{gra}$.  Material strength in small bodies provides a stabilizing effect against the host WD's tidal perturbation.  
        
        If $C$ is relatively large, this equation degenerates into 
        \begin{equation}
            d_\mathrm{str} = \Big(\dfrac{\sqrt{3}Gr^2\rho M_\star}{5k}\Big)^{1/3},
            \label{eq:dstru}
        \end{equation}
        which has the same format as the tidal disruption limit in the strength-dominated regime derived by \citet[][see their Eq.~(29)]{Brown2017}, but with slightly different coefficients.  The study of \cite{Brown2017} considers a pure tensile failure mode and neglects the internal pressure and shear stresses, while Eq.~(\ref{eq:tidallimit}) is derived from the global stress analyses and is applicable for various failure modes and strength levels.   
        
        \subsection{Maximum survivor size at different pericenter distances}
        
        According to Eq.~(\ref{eq:tidallimit}), $d_\mathrm{str}$ is size-dependent when the object's cohesive strength $C>0$ Pa.  In this case, the material strength and size of an object determine its integrity in the proximity of a WD.  With a given amount of cohesion, the maximum possible size of the survivors $r_\mathrm{max,sur}$ that can maintain their integrity at a certain pericenter distance $q$ can be estimated by this equation.

        \begin{figure}[htbp]
        \centering
        \includegraphics[width = 8.5 cm]{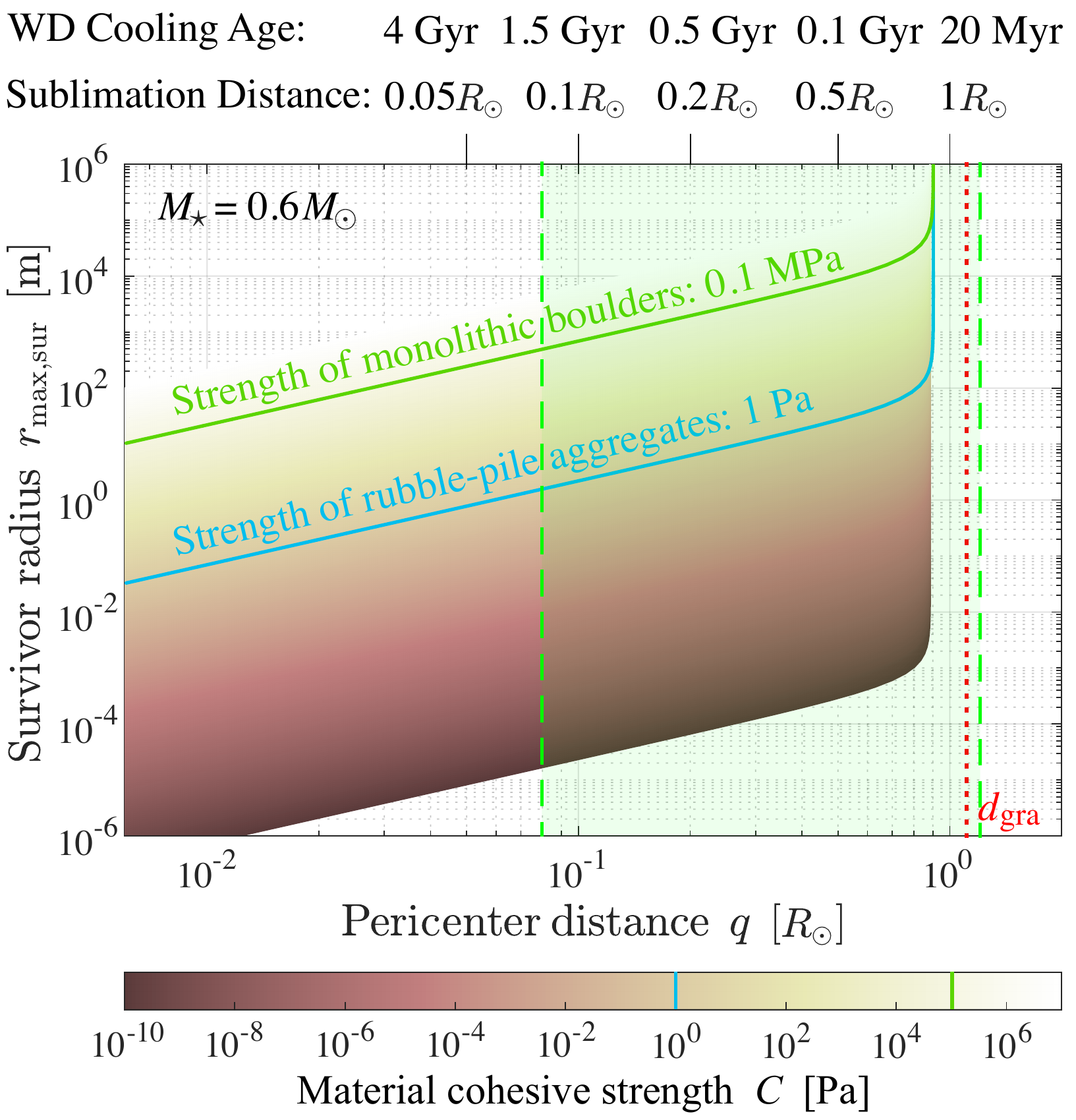}
        \caption{Maximum survivor size at different pericenter distances.  The colorbar represents the material strength magnitude.  The distances for silicate sublimation at various WD cooling ages are indicated on the top \citep[following Eq.~(1) given in][]{Rafikov2011}.  The greenish area in between the two green vertical dashed lines represent the radius range of tidal debris disks around WDs deduced from observations \citep[e.g.,][]{von_Hippel2007, Xu2017} and theoretical analyses \citep[e.g.,][]{Rafikov2012}.  The red vertical dotted line denotes the tidal disruption limit at the gravity-dominated regime.  The survivor sizes for material strength of 1 Pa and 0.1 MPa are highlighted by the blue and green solid curves, respectively.}
        \label{f:sizedistribution}
        \end{figure}
        
        Figure \ref{f:sizedistribution} shows the maximum possible size of debris around a WD with a typical mass $M_\star = 0.6M_\odot$ at different pericenter distances for a range of material cohesive strength.  A bulk density of $3000$ kg/m$^3$ and a friction angle of $30^\circ$ \citep[analogue to the properties of Solar System S-type asteroids;][]{Carry2012, Zhang2017} are used for these estimates.  Since $d_\mathrm{str}\propto M_\star^{1/3}$, this results can be scaled to different WD mass $M_\star$.
        
        Our evaluation of the tidal disruption limit $d_\mathrm{str}$ is consistent with the observed radius range of tidal debris disks around WDs (i.e., the greenish area in Fig.~\ref{f:sizedistribution}).  Within this range, grains and boulders that constitute the approached rubble-pile object are separated into individual debris if their connections are cohesionless.  Some meter-sized aggregates could survive if the van der Waals cohesive forces between fine grains \citep{Scheeres2010, Zhang2018} could provide cohesive strength on the order of 1 Pa.  With the typical material cohesive strength of meteorites \citep[i.e., 0.1--10 MPa; ][]{Pohl2020}, monolithic bodies with sizes up to 10 km can remain intact.  It is therefore appropriate to assume that the typical size of macroscopic tidal debris orbiting WDs ranges from $\sim$1 cm to $\sim$10 km.  In the following section, we will investigate the effect of WD magnetic fields on objects within this size range.   

\section{Orbital migration and circularization by Alfv{\'e}n-wave drag}
\label{sec:alfven}
After the tidal disruption, the fragments follow similar paths as their parent bodies.  Although some azimuthal dispersion occurs
after a few hundred orbital periods for the debris of parent bodies with semi-major axes similar to the main belt asteroids \citep{Malamud2020}, 
it takes much longer for the distant LPCs.  In an attempt to investigate the fragments' circularization process, we consider
their interaction with the magnetic fields of their host WDs.  

    \subsection{Orbital migration of tidal debris}
    
        When an object moves through a magnetized flowing plasma, it induces an impulse (i.e., Alfv{\'e}n wave) traveling along the magnetic field that generates significant drag effects on its motion \citep{Drell1965}.  Here we consider a tidal fragment orbiting a magnetic WD on a highly eccentric orbit with a pericenter within the tidal disruption limit of the WD.  The radius and density of this fragment are denoted as $r$ and $\rho$, respectively, and thus the mass $m = 4\pi r^3\rho/3$ (assumed to be spherical).  
        
        Assuming the WD has a surface magnetic flux density $B_\star$, for an object at a distance $d$ from the WD, where the flux density $B=B_\star(R_\star / d)^3$, the Alfv{\'e}n wave is emitted along a wing with a power \citep{Drell1965, Neubauer1980}
        \begin{equation}
            P \simeq \pi U^2/R_{\rm wing},
        \end{equation}
        where $U=rvB$ is the electric potential of the electric field induced by this object across its radius $r$ 
        (with an orbital speed $v$), and $R_{\rm wing} = \mu_0 \bar{v}_\mathrm{A} = \mu_0 v_\mathrm{A}\sqrt{1+M_\mathrm{A}^2+2M_\mathrm{A}\sin{\theta}}$ is the effective resistance provided by the surrounding plasma. The permeability of free space $\mu_0 = 4\pi\times10^{-7}$ N A$^{-2}$.  The Alfv{\'e}n speed, $v_\mathrm{A} = B/\sqrt{\mu_0\rho_\mathrm{g}}$ (where $\rho_\mathrm{g}$ is the ionized gas mass density), and the Alfv{\'e}n Mach number $M_\mathrm{A}=v/v_\mathrm{A}$.  We neglect the spin of the WD such that $v$ is assumed to be the orbital motion of the object.
        \footnote{With the model parameters we adopt below (i.e., $q=10^{-3}$ au and $M_\star =0.6 M_\odot$; see Table \ref{t:parameter}), the orbital 
        frequency at the pericenter corresponds to a period of $2 \pi (2 q^3/GM_\star)^{1/2} \sim 600$ s, which is much shorter 
        than the spin periods of most WDs \citep[e.g.,][]{koester1998, Hermes2017}.}  $\theta$ is the deviation of the object's velocity from being perpendicular to the magnetic field.

        \subsubsection{Energy dissipation rate}
        Interacting with the Alfv{\'e}n-wave drag, the body's kinetic energy per unit mass is dissipated at a rate
        \begin{equation}
        \label{eq:speedrate}
            \dfrac{1}{2}\dfrac{\mathrm{d}v^2}{\mathrm{d}t} = -\dfrac{P}{m} = -\dfrac{3 v^2 B^2}{4 \rho r\mu_0 \bar{v}_\mathrm{A}}
        \end{equation}
        This Alfv{\'e}n-wing drag force has a similar, but not identical, dependence on the stellar magnetic field 
        as the diamagnetic planetary debris model, which is based on the assumption of a back reaction by the distorted field lines wrapped around the moving object \citep{King1993, Hogg2021}.  These two assumptions lead to a magnitude difference in the drag force 
        by a factor $v/\bar{v}_\mathrm{A}$.  Since $\bar{v}_\mathrm{A} \propto B$, this factor can depart from unity in the limit of modest $B_\star$.
        
        As the magnetic flux density $B$ decay rapidly with a larger distance $d$, the orbital modifications caused by this mechanism is just effective around the pericenter, where the distance $d=q=a(1-e)$, where $e$ and $a$ are the orbital eccentricity and semi-major axis, respectively.  The total energy change during one passage can be estimated by
        \begin{equation}
            \Delta E = \dfrac{1}{2}\dfrac{\mathrm{d}v^2}{\mathrm{d}t}\Delta t_\mathrm{peri}.
        \end{equation}
        The duration of the pericenter passage $\Delta t_\mathrm{peri} = \alpha/\Omega$, where the angular speed at pericenter $\Omega = \sqrt{GM_\star a(1-e^2)}/a^2(1-e)^2$, and $\alpha$ is the pericenter passage angle within which the Alfv{\'e}n-wave drag takes effect.  Given that $1-e = \varepsilon \ll 1$, and $\Omega = \sqrt{GM_\star /a^3\varepsilon^3}$, the energy change can be given as
        \begin{equation}
            \Delta E = -\dfrac{3\alpha v^2 B_\star ^2R_\star ^6}{4 \rho r \mu_0 \bar{v}_\mathrm{A}a^5\varepsilon^5}\sqrt{\dfrac{a\varepsilon}{GM_\star }}.
        \end{equation}
        Substituting $v$ with the speed at perihelion $v_\mathrm{peri} = \sqrt{GM_\star (2-\varepsilon)/a\varepsilon} \approx \sqrt{2GM_\star /a\varepsilon}$, 
        \begin{equation}
            \Delta E = -\dfrac{3\alpha B_\star ^2R_\star ^6}{2 \rho r \mu_0 \bar{v}_\mathrm{A}a^5\varepsilon^5}\sqrt{\dfrac{GM_\star }{a\varepsilon}}.
        \end{equation}
        With the total specific orbital energy $E = -GM_\star /2a$, the energy change rate per orbit can be estimated by
        \begin{equation}
        \label{eq:energyrate}
            \dfrac{\Delta E}{E} = \dfrac{3\alpha B_\star ^2R_\star ^6}{\rho r \mu_0 \bar{v}_\mathrm{A}a^4\varepsilon^5GM_\star}\sqrt{\dfrac{GM_\star }{a\varepsilon}}.
        \end{equation}

    \subsubsection{Semi-major axis change rate}
    
        The energy dissipation would lead to semi-major axis decay, which can be estimated by
        \begin{equation}
            \label{eq:arate}
            \dfrac{\Delta a}{a} = -\dfrac{\Delta E}{E}
        \end{equation}
        
    \subsubsection{Eccentricity change rate}
    \label{sec:alfven:e}
        
        With the assumption that this effect is only effective near the pericenter, the pericenter distance $q$ can be regarded as a constant, $q=q_0$, and therefore
        \begin{equation}
            \Delta e = -(1-e)\dfrac{\Delta E}{E},
        \end{equation}
        
    \subsection{Orbital circularization timescale}
        \label{sec:alfven:timescale}
        
        Now we investigate the long-term orbital migration of an object due to the Alfv{\'e}n-wave drag effect.  The initial orbital semi-major axis and eccentricity are denoted as $a_0$ and $e_0$, respectively.  After $n$ times pericenter approaches, the semi-major axis $a_n$ is described as the following polynomial recursion format,
        \begin{equation}
            a_n = a_{n-1} + \Delta a_{n-1} = a_{n-1} - \lambda a_{n-1}^2,
        \end{equation}
        where $\lambda = \Delta E_\mathrm{n-1}/(E_\mathrm{n-1}a_\mathrm{n-1}) = \Delta E_0/(E_0a_0) = \mathrm{constant}$.  
        
        The analytical solution of this recursion is given by 
        \begin{equation}
            a_n = a_0\mathbf{I}\mathbf{T}^n \mathbf{E}_0,
        \end{equation}
        where $\mathbf{T} \in \mathbb{Z}^{2^n \times 2^n}$ is the upper triangular transfer matrix with elements
        \begin{equation}
            T_{jk} = (-1)^{k-j} {k \choose {k-j}}.
        \end{equation}
        The vector $\mathbf{I}=[\delta_{i1}]_{i=1,2,...,2^n}\in \mathbb{Z}^{1\times2^n}$, where $\delta_{i1}$ is the Kronecker symbol.  The vector $\mathbf{E}_0=[(\Delta E_0/E_0)^{i-1}]_{i=1,2,...,2^n} \in \mathbb{Z}^{2^n\times1}$.  
    
        The semi-major axis after $n$ encounters, $a_n$, can be expressed more concisely in such a form,
        \begin{equation}
            \label{eq:an}
            a_n = \sum_{i=0}^{2^n-1} (-1)^{i} a_0 P_{i}(n)\bigg(\dfrac{\Delta E_0}{E_0}\bigg)^{i},
        \end{equation}
        where $P_{i}(n)$ is the $i$-th degree polynomials of $n$.
        
        Given that the corresponding orbital eccentricity $e_n = 1-{q}/{a_n}$, the number of orbits required to reduce the eccentricity to 0.1, $n_{e = 0.1}$, can be obtained by solving $a_{n_{e = 0.1}} = 10a_0(1-e_0)/9$.
        By substituting Eq.~(\ref{eq:an}) into the above equation, we find that, approximately, $n_{e = 0.1}$ is proportional to $(\Delta E_0/ E_0)^{-1}$ and slowly decreases with a larger $e_0$ for a given pericenter distance.
    
        The total timescale is then given as (assuming the starting point is the apocenter)
        \begin{equation}
            \Delta T_{e = 0.1} = \pi\sqrt{\dfrac{ a_0^3}{G M_\star}} + \dfrac{2\pi}{\sqrt{G M_\star}}\sum_{i=1}^{n_{e = 0.1}} a_i^{1.5}.
        \end{equation}
        In the case where $n_{e = 0.1}\gg 1$, $\Delta T_{e = 0.1}$ is proportional to the total number of orbits $ n_{e = 0.5}$, i.e., 
        \begin{equation}
        \label{eq:timescale_slow}
            \Delta T_{e = 0.1} \propto \sqrt{\dfrac{ a_0^3}{G M_\star}} \bigg(\dfrac{\Delta E_0} {E_0}\bigg)^{-1} = \dfrac{\rho r \mu_0 \bar{v}_\mathrm{A} q_0^{5.5} \sqrt{a_0}}{3\alpha  B_\star ^2 R_\star ^6}.
        \end{equation}
        In another extreme where the orbital circularization is fast and $n_{e = 0.1}$ is small, $\Delta T_{e = 0.1}$ mainly depends on the initial orbital semi-major axis, i.e.,
        \begin{equation}
        \label{eq:timescale_fast}
            \Delta T_{e = 0.1} \propto \sqrt{\dfrac{ a_0^3}{G M_\star}}.
        \end{equation}
        Therefore, in general, we have
        \begin{equation}
        \label{eq:timescale}
            \Delta T_{e = 0.1} \propto \sqrt{\dfrac{ a_0^3}{G M_\star}} \bigg(\dfrac{\Delta E_0} {E_0}\bigg)^{-1/\beta},
        \end{equation}
        where $\beta \gtrsim1 \in \mathbb{R}$ and depends on the circularization efficiency.  When the efficiency is low, $\beta \rightarrow 1$ and Eq.~(\ref{eq:timescale}) degenerates into Eq.~(\ref{eq:timescale_slow}).  On the other hand, if the object's orbit can be circularized in a few pericenter approaches, $\beta \rightarrow \infty$ and Eq.~(\ref{eq:timescale}) degenerates into Eq.~(\ref{eq:timescale_fast}).
        
        To give some quantitative estimates on the timescale, we consider orbital migration of an object around a typical WD, whose $M_\star = 0.6 M_\odot$, $R_\star \approx 0.0125R_\odot$ \footnote{The WD radius is estimated by $R_\star = 0.0127R_\odot (M_\star / M_\odot)^{-1/3} [1-0.607(M_\star / M_\odot)^{4/3}]^{1/2}$ \citep{Veras2020}.}. Assuming that $B_\star = 100$ T (i.e., $10^6$ G in cgs units), $\rho \sim 3000$ kg/m$^3$, $r=1$ m, $\bar{v}_\mathrm{A}=3\times10^7$ m/s \footnote{The Alfv{\'e}n wave drag is stronger with a smaller $\bar{v}_\mathrm{A}$. For a conservative estimate, we adopt a constant large value for $\bar{v}_\mathrm{A}$ in this study. The actual value of $\bar{v}_\mathrm{A}$ depends on the magnetic flux density, the surrounding ionized gas mass density, the orbital motion of the encountered object, and the magnetic field direction, which are poorly constrained based on current observations and whose effects will be examined for future studies.}, $a_0=10$ au, $\varepsilon_0=10^{-4}$ \footnote{So the perihelion distance $q\approx 0.2R_\odot$, which is within the tidal disruption limit of most WDs; see Fig.~\ref{f:sizedistribution}.}, and $\alpha$ is set to a constant $\pi/2.24$ by calibration with numerical simulations in Section \ref{sec:numerical}, the energy dissipation rate $\Delta E/E \sim 0.027$. In this particular case, the object's orbital eccentricity decays to 0.1 after $\sim$306,000 orbits, which corresponds to a timescale about 2.7 kyr. In general, the timescale scales as  $2.7(r/1~\mathrm{m})(a_0/10~\mathrm{au})^{0.5}$ kyr approximately, suggesting that the Alfv{\'e}n-wave drag is an efficient mechanism for the orbital circularization of sub-100 km-sized planetary remnants around magnetic WDs.  
        
        With these model parameters, $v/{\bar v}_{\rm A} \sim 0.032-0.043$ such that the Alfv{\'en} wave drag
        force we used here is an order of magnitude smaller than that adopted by \citet{Hogg2021} for the diamagnetic 
        planetary debris model. Despite our conservative choices, the high efficiency of circularization by this process
        appears to be robust.
        The measured $B_\star$ for WDs has a wide range \citep[$10^{3-9}$ G; ][]{Ferrario2015}.  Since the Alfv{\'e}n speed  
        $v_{\rm A} \propto B$, $\Delta E/E \propto B$ and $dv/dt \propto B$ (Eqs. (\ref{eq:energyrate}) \& (\ref{eq:dvdt})). 
        The circularization timescale is proportional to $B_\star ^{-1}$.  Nevertheless, tidal fragments of km-sized LPCs (with $a_0 \sim 10^3$ au) can be circularized in $\lesssim 1$ Gyr around weakly magnetized WDs (with kilogauss $B_\star$).

    \begin{deluxetable}{ccl}
    \tabletypesize{\scriptsize}
    \tablecolumns{3}
    \tablewidth{0pt}
    \tablecaption{Parameter setup in the numerical simulations for testing the effect of Alfv{\'e}n-wave drag.
    \label{t:parameter}}
    \tablehead{
        \colhead{Symbol} & \colhead{Value} & \colhead{Meaning}
        }
        \startdata
        $M_\star$ & 0.6 $M_\sun$ &  WD mass \\
        $R_\star$ & 0.0125 $R_\sun$ & WD radius \\
        $B_\star$ & 100 T (or $10^6$ G) & Magnetic field at the surface of WD \\
        $\bar{v}_\mathrm{A}$ & $3\times 10^7$ m/s & Scaled Alfv{\'e}n speed \\
        $r$ & 0.01--10$^4$ m & Debris radius\\ 
        $a_0$ & 1--10$^4$ au & Debris semi-major axis\\ 
        $q_0$ & 0.001--0.002 au & Debris periastron distance \\
        $\varepsilon_0$ & $10^{-7}$--$10^{-3}$ & $1-e_0$ \\
        $\rho$ & 3000 kg/m$^3$ & Debris bulk density  \\
        $\alpha$ & $\pi/2.24$ & Empirical pericenter passage angle \\
        \enddata
    \end{deluxetable}
        
\section{Numerical simulations}
\label{sec:numerical}

    \begin{figure}
    \centering
    \includegraphics[width = \linewidth]{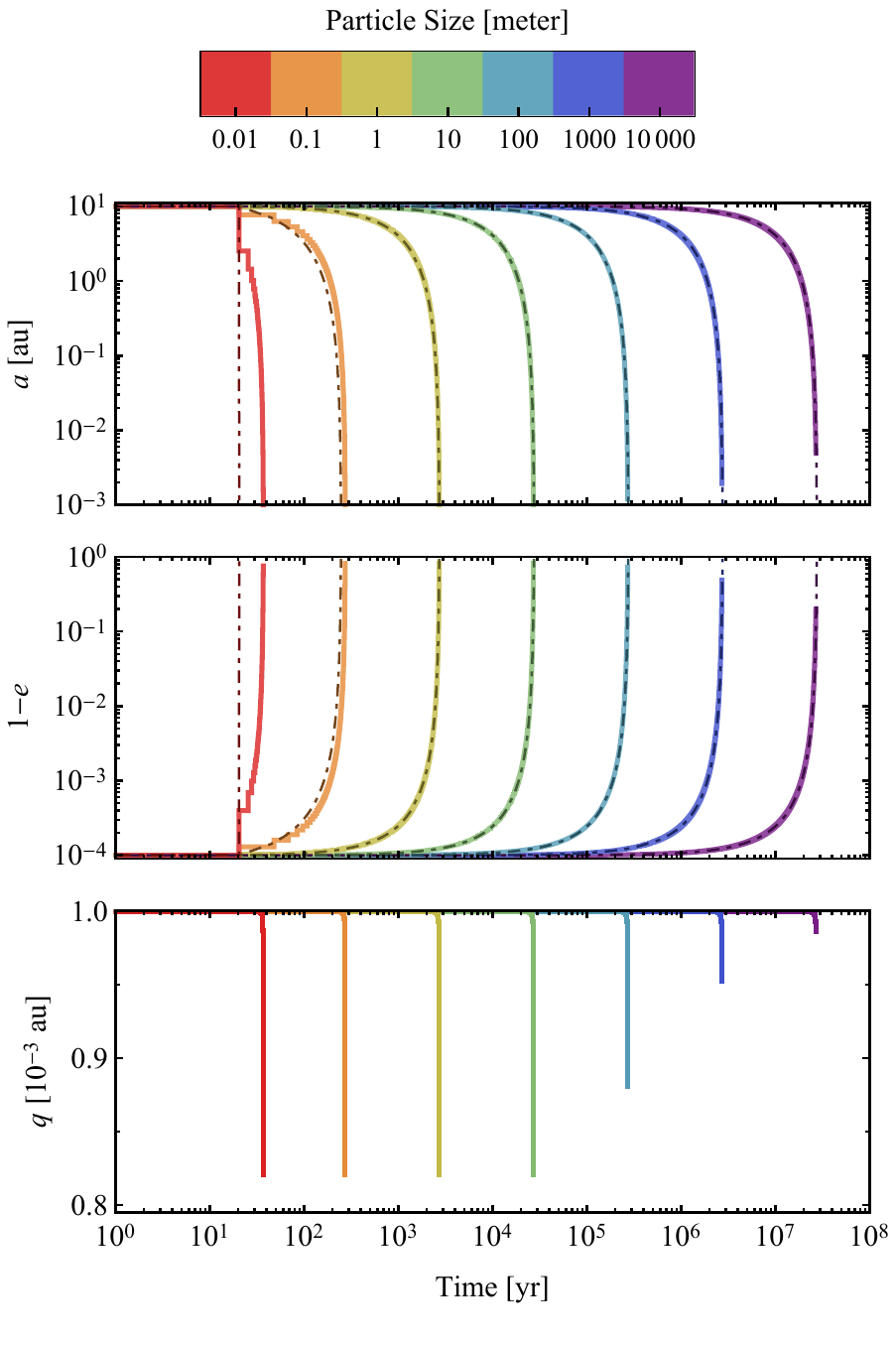}
    \caption{Orbital circularization of different sized fragments caused by the Alfv{\'e}n-wave drag from an initial orbit with $a_0 = 10$ au and $e_0 = 0.9999$. Evolution of orbital semi-major axis $a$, eccentricity represented by $1-e$, and the periastron distance $q$ are plotted. Lines are color-coded with sizes of fragments. Solid lines are results from numerical integrations, while dot-dashed lines are analytic estimations (Section \ref{sec:alfven:timescale}).}
    \label{f:sim_set1}
    \end{figure}

    In this section, we validate the analytical estimation presented in Section \ref{sec:alfven} by numerical simulations. We use the the IAS15 integrator \citep{Rein2015} within the REBOUND package \citep{Rein2012} to numerically integrate orbits of the tidal debris and determine the orbital migration timescales on various initial orbits. The debris is treated as a test particle, which is subject to the Alfv{\'e}n-wave drag in addition to the gravitational force from the host. The Alfv{\'e}n-wave drag is opposite to the particle's motion and the magnitude of deceleration at a distance $d$ from the WD can be characterized by
    \begin{equation}
        \dfrac{\mathrm{d}v}{\mathrm{d}t} = -\dfrac{ 3v B_\star^2 R_\star^6} {4\rho r d^6 \mu_0 \bar{v}_\mathrm{A}}.
\label{eq:dvdt}
    \end{equation}  

    \begin{figure*}
    \centering
    \includegraphics[width = \linewidth]{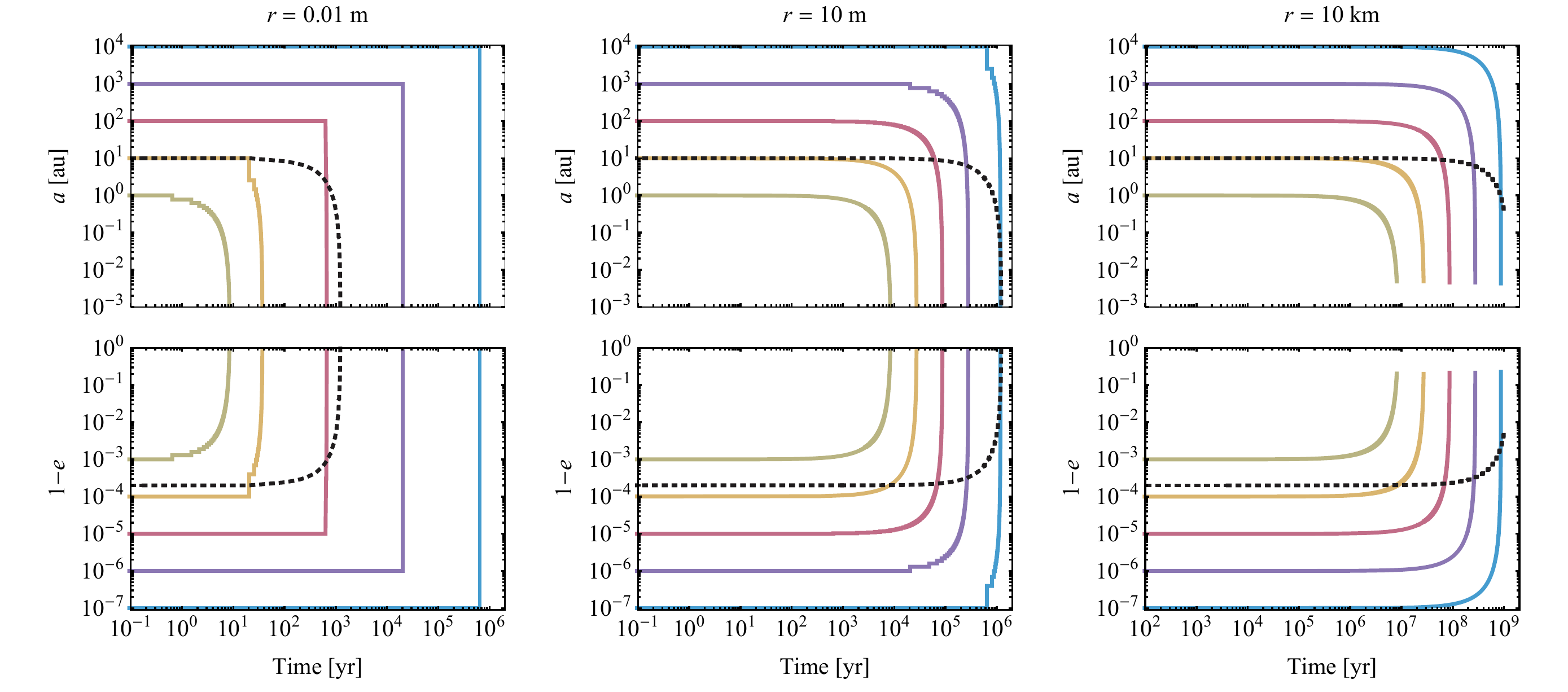}
    \caption{Orbital circularization of fragments with sizes at $r=1$ cm (left column), $r=10$ m (middle column) and $r=10$ km (right column) caused by the Alfv{\'e}n-wave drag. Evolution of orbital semi-major axis $a$, eccentricity represented by $1-e$ are plotted. Colored solid lines are cases with an initial periastron distance $q_0 = 10^{-3}$ au, while black dashed lines are cases with an initial perastron distance $q_0=2\times10^{-3}$ au.}
    \label{f:sim_set2}
    \end{figure*}
    
    We consider a typical WD with a mass $M_\star = 0.6 M_\odot$ and a radius $R_\star=0.0125R_\odot$. Other parameters are summarized in Table \ref{t:parameter}.  For all cases, we start numerical integrations from the apocenter of an initial orbit with a semi-major axis $a_0$ and an eccentricity $e_0$.
    
    In the first test, we adopt the initial orbit with $a_0=10$ au and $e_0=0.9999$ to examine the effect of debris size and make comparisons with the analytic approximation in Section \ref{sec:alfven}. Figure \ref{f:sim_set1} shows the orbital evolution of the tidal debris with sizes ranging from 0.01 m to 10 km. The Alfv{\'e}n-wave drag is only effective at the proximity of the WD as $\mathrm{d}v/\mathrm{d}t \sim -d^{-6}$, so it gets intensified rapidly near the periastron and is negligible for the rest of the time. In addition, since the magnetic drag force is inversely proportional to the size of debris, a smaller fragment feels a much stronger drag effect than a bigger one. As a result, a small fragment ($r=0.01\;\mathrm{m}$) could lose a significant fraction of its orbital energy, e.g., $\Delta E / E \gtrsim 1/2$, and its orbit is rapidly circularized within a few periastron passages. In Figure \ref{f:sim_set1}, the staircase-like variations of $a$ and $1-e$ are evident for the smallest fragment ($r=0.01$ m) during the first few orbits. We stop the numerical integration when $a$ becomes smaller than $10^{-3}$ au, which equals to $q_0$, the initial periastron distance. Further orbital shrinkage would require proper treatment of general relativity, which is not the focus of this paper. For fragments with radii larger than 100 m, our numerical integrations end at slightly larger semi-major axes due to time steps being extremely small. 

    As noted in Section \ref{sec:alfven}, the circularization timescale is proportional to the size of debris. We overplot the analytical estimation for each sizes in thin dot-dashed lines in the $a$ and $1-e$ plots for comparison. The numerical results show good consistency except for the smallest size ($r=0.01$ m) where the analytical result underestimates the circularization timescale because the orbital circularization is fast and the timescale is determined by initial orbital period (see Eq.~(\ref{eq:timescale_fast})). Numerical results also suggest that, during the orbital circularization, the periastron distance remains almost unchanged in all cases, validating the assumption adopted for the analytical estimation (i.e., see Section \ref{sec:alfven:e}).

    We further explore the dependence of orbital circularization on the initial locations of debris for given fragment sizes ($r = 0.01 \;\mathrm{m}$, $r = 10 \;\mathrm{m}$ and $r = 10 \;\mathrm{km}$). Figure \ref{f:sim_set2} presents the evolution of $a$ and $1-e$ as a function of time for each considered size of debris. We keep the periastron distance $q_0=10^{-3} \;\mathrm{au}$ constant and vary the initial semi-major axis $a_0$ from $1\;\mathrm{au}$ to $10^4\;\mathrm{au}$ (in colored solid lines) except for one initial orbit with $a_0=10 \;\mathrm{au}$ and $q_0=2\times 10^{-3} \;\mathrm{au}$ (in black dashed line). For debris at size $r=0.01\;\mathrm{m}$, the Alfv{\'e}n-wave drag is so efficient that it can circularize the orbit of fragments from the Oort cloud after one passage with $q_0=10^{-3}\;\mathrm{au}$ (consistent with the estimation of Eq.~(\ref{eq:timescale_fast})). For debris at size $r=10\;\mathrm{m}$ and $r=10\;\mathrm{km}$, the orbital circularization timescale is consistent with the analytical estimation given by Eq.~(\ref{eq:timescale_slow}), i.e., $\sim$1 Myr for $r=10\;\mathrm{m}$ and $\sim$1 Gyr for $r=10\;\mathrm{km}$ for debris from the Oort cloud.

    Our analytical and numerical investigations reveal that the Alfv{\'e}n drag-assisted circularization is a rapid and robust process that can completely circularize large residual planetesimals from the Oort cloud around magnetic WDs.  The derived time scale is comparable to or shorter than other rapid circularization methods proposed by previous theoretical works that consider the effect of dust or gas drag of a pre-existing gaseous or massive compact disk \citep{Grishin2019, Malamud2021} or tidal migration of planetesimals with high internal viscosity \citep{OConnor2020}. Moreover, the Alfv{\'e}n drag mechanism does not require the ram-pressure drag by hypothetical gaseous or debris disks or by tidal dissipation of bodies with molten interiors.

\section{Metamorphsis from LPCs to close-in fragments}
\label{sec:lpctodebris}
In Section \ref{sec:lpcs}, we adopted the assumption that WDs bear rich reservoirs of residual planetesimals 
at distances comparable to the Oort cloud.  
This population of residual planetesimals can sustain protracted injection of LPCs into their 
tidal disruption radius close to their host stars.  When their pericenter distances are reduced below $d_{\rm gra}$ 
and $d_{\rm str}$, the residual planetesimals are tidally disrupted (see Section \ref{sec:materialstrength}).  
According to Fig.~\ref{f:sizedistribution} and Eq.~(\ref{eq:tidallimit}), at the pericenter distance $q=10^{-3}$ au, the maximum sizes of the surviving fragments $r_{\rm max, sur}$ is $\sim$1 m and $\sim$10 km 
for LPCs with rubble-pile and monolithic structures, respectively.  These fragments approximately follow the paths of 
their LPC parent bodies.  Their orbits must be circularized before they 
can settle into debris disks interior to $d_{\rm gra} (\sim R_\odot)$.  Such disks are commonly found
around young ($<$ Gyr old) WDs with signatures of metal enrichment through recent accretion of refractory material
\citep{mullally2007, farihi2009, Farihi2016, Manser2020}.
This requirement is challenging since the dynamical time scale for LPCs with $a \sim 10^{3-4}$ au is 
$P \sim 10^{5-6}$ yr.  Around some recently emerged WDs (with ages of $\lesssim 10^8$ yr), the 
LPCs' fragments must be effectively circularized within $\sim 10^{2-3}$ orbits.  

As discussed in the beginning of this paper (e.g., see Fig.~\ref{f:schematic}), one efficient mechanism to transform the residual planetesimals' nearly parabolic to nearly circular 
orbits is through their interaction with their magnetized host WDs during their pericenter 
passages.  Our analytic and numerical results (Sections \ref{sec:alfven} 
\& \ref{sec:numerical}) show that the circularization time scale $\Delta T_{e = 0.1} \propto q_0^{5.5}$ for a constant $a_0$ (Eq.~(\ref{eq:timescale_slow})).  
We consider modal parameters for LPCs that enters inside the tidal disruption radius during their 
pericenter approach.  For a fiducial residual planetesimal with $r = 10$ km, $q=10^{-3}$ au, 
and $a = 10^4$ au, $\Delta T_{e = 0.1} < 1$ Gyr.
The results in Section \ref{sec:numerical} show rapid decreases in $a$ and $e$ but little change in $q$.
For the same pericenter distance, $\Delta T_{e = 0.1} \propto r a^{1/2}$, which decreases during the tidal downsizing 
and orbital decay processes.  Therefore, the timescale for typical LPCs to circularize their orbits is a 
small fraction of that for them to be transformed from residual planetesimals in the Oort cloud.  

After the eccentricity of individual fragments is reduced to negligible values, they continue to undergo orbital decay 
at a location $a$ with a rate (derived from Eqs.~(\ref{eq:energyrate}) \& (\ref{eq:arate}))
\begin{equation}
    {\dot a} = {\Delta E \over 2 \pi E} \left( { G M \over a} \right)^{1/2}.
\end{equation}
The characteristic orbital decay timescale,
\begin{equation}
    \tau_{\rm a} = {a \over {\dot a}} = { 4 \pi \rho_{\rm frag} r_{\rm frag } v_{\rm A} \mu_0 a^5 \over 3 \alpha B_\ast^2 R_\ast^6},
\end{equation}
increases with both the fragment's size $r_{\rm frag}$, density $\rho_{\rm frag }$, and semi-major axis $a$.  

During the fragments' orbit decay,  the number of LPCs being replenished inside the tidal disruption radius can be estimated by,
\begin{equation}     
N_{\rm LPC} \simeq \dot{n}_{\rm LPC} \tau_{\rm a}.
\end{equation}
In the limit $N_{\rm LPC} \leq 1$, single disks
would form from the debris of a LPC and decay before the arrival of another intruding LPC, such that
$N_{\rm LPC}$ is equivalent to the duty cycle. In the limit $N_{\rm LPC} > 1$, multiple inclined disks 
coexist, since the fragments of each LPC form a separate plane.  This possibility may account for the
coexistence of two or more rings around a 3 Gyr WD \citep{debes2019}. 

As the fragments undergo orbital decay and migrate inwards, the WD tidal force intensifies and the largest 
surviving fragments continue to be downsized $r_{\rm max, sur} \propto q^{3/2}$ (see Fig.~\ref{f:sizedistribution}
and Eq.~(\ref{eq:tidallimit})). In a complementary study, \citet{Hogg2021} showed that under the influence 
of WDs' magnetic field, small fragments in the proximity of WDs may be cleared on timescale much shorter than that for the circularization 
of LPCs' orbits.  It is not clear whether the distorted fields lines or overlapping Alfv{\'e}n wings
may interfere with each other in debris disks where the fragments contribute to a significant area filling factor.  
The efficiency of this magnetosphere-debris disk interaction remains an outstanding issue.

\section{Summary}
\label{sec:summary}

We adopt a conventional scenario that the heavy-element enriched surfaces of many WDs is a manifestation of the
continuous accretion of the relic building blocks from the planetary-assemblage era. Since the inferred delivery 
rates of metals onto these WDs is comparable to the cometary bombardment flux onto the Sun, we use
the Solar System as an analogue to ascribe the parent bodies of these pollutants to be LPCs originated
from a population of residual planetesimals at locations comparable to the Oort cloud (located at $\sim 10^{3-4}$ au).  

We address several key issues concerning this scenario: 

\begin{enumerate}
    \item the profusion of distant residual planetesimals,
    \item the transformation of residual planetesimals into LPCs, 
    \item  the tidal disruption of LPCs, 
    \item the formation of close-in disks from the circularized fragments, and 
    \item the accretion flow from dusty disks onto WDs.
\end{enumerate}

For the origin of the residual planetesimals, we adopt the hypothesis that they form in the proximity of 
giant planets and are launched by the planetary perturbation to large distances from their host stars.  
We briefly recapitulate (Section \ref{sec:lpcs}) the subsequent pathways of their dynamical 
evolution as their host stars age.  For discussions on the dominant physical effects, it is helpful to categorize the chronology into epochs of:
\begin{enumerate}
    \item planet formation,
    \item natal gaseous depletion and planetesimal scattering, 
    \item dispersal of cohort-star clusters,
    \item secular evolution due to Galactic tides, passing GMCs and stars,
    \item host stars' asymptotic-giant and mass-loss phase, and
    \item WD cooling sequences.
\end{enumerate}

The pivotal physical effects include: 
\begin{enumerate}
    \item planetary scattering due to close encounters, resonant and secular perturbation;
    \item external perturbation due to Galatic tides, passing stars and GMCs; 
    \item dispersal of volatile components and orbital expansion during the host stars' post-main-sequence evolution;
    \item planetesimals' material strength and internal composition; and 
    \item host WD luminosity, tidal and magnetic fields. 
\end{enumerate}

The main assumption of our scenario is that giant planets are common around WDs 
and their main-sequence progenitors. This assumption is tenable since such planets are 
common around stars more massive than the Sun, including WD progenitors.  
These planets' contribution is essential for the relocation of a rich
population of residual planetesimals from their neighborhood to large distances from their host stars,
analogous to the Oort cloud in the Solar System.  At the outlying regions of host stars' gravitational domain, 
external perturbation can lead to a continuous injection of LPCs over Gyr timescales.  Although such giant 
planets may also slow down and temporarily interrupt the gradual decrease of residual planetesimals' pericenter distances and 
their conversion into LPCs, this planetary barrier is dynamically permeable and LPCs can effectively bypass it 
as shown in the context of the Solar System.

We show (Section \ref{sec:materialstrength}) that LPCs on nearly parabolic orbits can undergo tidal 
disruption provided their pericenter distances are within a fraction of the solar radius.  However,
these  fragments generally follow the orbits of their LPC parent bodies.  In order for them to be accreted onto 
the compact WDs, their semi-major axis must be greatly reduced through both angular momentum and energy loss.
We show (Section \ref{sec:alfven}) that around magnetized WDs, the fragments excite Alfv{\'e}n waves during their 
closest approach and the resulting magnetic torque induces them to undergo orbital decay and circularization.  
Based on our analytic results and numerical confirmation (Section \ref{sec:numerical}), we suggest this magnetic 
interaction is a promising process for delivering heavy elements to the proximity of the WDs.  
 
This magnetic interaction may further remove angular momentum and energy from the fragments after their orbits 
are circularized.  In addition to magnetic breaking, the last leg of their journey to WD photospheres involves 
many additional physical processes including the Yarkovsky and YORP effects, PR drag, rotational and tidal downsizing,
sublimation, collisional break-ups of fragments, 
viscous diffusion, and magnetosphere-disk interaction\citep{Rafikov2011, Veras2014, Veras2015a, Hogg2021}.  
In an effort to model the IR excess
around metal-enriched WDs, we plan to address these issues in a subsequent investigation.

\begin{acknowledgments}
The authors thank an anonymous referee for careful reading of the manuscript and helpful suggestions.  We also thank
Roman Rafikov, Dimitri Veras, Xiaochen Zheng, and Simon Portegies Zwart for useful discussions and comments. S.-F.L. acknowledges support from National Natural Science Foundation of China under grant number 11903089 and from Guangdong Basic and Applied Basic Research Foundation under grant number 2021B1515020090 and 2019B030302001. Y.Z. acknowledges funding from the Universit\'e C\^ote d'Azur ``Individual grants for young researchers'' program of IDEX JEDI and the European Union's Horizon 2020 research and innovation program under grant agreement No. 870377 (project NEO-MAPP).

\end{acknowledgments}

\software{REBOUND \citep{Rein2012}, REBOUNDx \citep{Tamayo2020}}

\bibliographystyle{aasjournal}
\bibliography{references}{}

\begin{thebibliography}{}
\expandafter\ifx\csname natexlab\endcsname\relax\def\natexlab#1{#1}\fi
\providecommand{\url}[1]{\href{#1}{#1}}

\bibitem[{{Adams}(2010)}]{adams2010}
{Adams}, F.~C. 2010, \araa, 48, 47

\bibitem[{Adams {et~al.}(2013)Adams, Anderson, \& Bloch}]{Adams2013}
Adams, F.~C., Anderson, K.~R., \& Bloch, A.~M. 2013, Monthly Notices of the
  Royal Astronomical Society, 432, 438.
\newblock \url{https://doi.org/10.1093/mnras/stt479}

\bibitem[{{Andrews} {et~al.}(2018){Andrews}, {Huang}, {P{\'e}rez}, {Isella},
  {Dullemond}, {Kurtovic}, {Guzm{\'a}n}, {Carpenter}, {Wilner}, {Zhang}, {Zhu},
  {Birnstiel}, {Bai}, {Benisty}, {Hughes}, {{\"O}berg}, \&
  {Ricci}}]{andrews2018}
{Andrews}, S.~M., {Huang}, J., {P{\'e}rez}, L.~M., {et~al.} 2018, \apjl, 869,
  L41

\bibitem[{Barber {et~al.}(2012)Barber, Patterson, Kilic, Leggett, Dufour,
  Bloom, \& Starr}]{Barber2012}
Barber, S.~D., Patterson, A.~J., Kilic, M., {et~al.} 2012, The Astrophysical
  Journal, 760, 26.
\newblock \url{https://doi.org/10.1088/0004-637x/760/1/26}

\bibitem[{Bareither {et~al.}(2008)Bareither, Edil, Benson, \&
  Mickelson}]{Bareither2008}
Bareither, C.~A., Edil, T.~B., Benson, C.~H., \& Mickelson, D.~M. 2008, Journal
  of geotechnical and geoenvironmental engineering, 134, 1476

\bibitem[{{Barrientos} \& {Chanam{\'e}}(2021)}]{Barrientos2021}
{Barrientos}, M., \& {Chanam{\'e}}, J. 2021, arXiv e-prints, arXiv:2102.07790

\bibitem[{{Boe} {et~al.}(2019){Boe}, {Jedicke}, {Meech}, {Wiegert}, {Weryk},
  {Chambers}, {Denneau}, {Kaiser}, {Kudritzki}, {Magnier}, {Wainscoat}, \&
  {Waters}}]{boe2019}
{Boe}, B., {Jedicke}, R., {Meech}, K.~J., {et~al.} 2019, \icarus, 333, 252

\bibitem[{Bromley \& Kenyon(2019)}]{Bromley2019}
Bromley, B.~C., \& Kenyon, S.~J. 2019, The Astrophysical Journal, 876, 17.
\newblock \url{https://doi.org/10.3847/1538-4357/ab12e9}

\bibitem[{Brown {et~al.}(2017)Brown, Veras, \& Gänsicke}]{Brown2017}
Brown, J.~C., Veras, D., \& Gänsicke, B.~T. 2017, Monthly Notices of the Royal
  Astronomical Society, 468, 1575.
\newblock \url{https://doi.org/10.1093/mnras/stx428}

\bibitem[{Carry(2012)}]{Carry2012}
Carry, B. 2012, Planetary and Space Science, 73, 98 , solar System science
  before and after Gaia.
\newblock
  \url{http://www.sciencedirect.com/science/article/pii/S0032063312000773}

\bibitem[{{Chen} {et~al.}(2019){Chen}, {Zhou}, {Xie}, {Yang}, {Zhang}, {Liu},
  {Liang}, {Yu}, \& {Yang}}]{Chen2019}
{Chen}, D.-C., {Zhou}, J.-L., {Xie}, J.-W., {et~al.} 2019, Nature Astronomy, 3,
  69

\bibitem[{{Cumming} {et~al.}(2008){Cumming}, {Butler}, {Marcy}, {Vogt},
  {Wright}, \& {Fischer}}]{cumming2008}
{Cumming}, A., {Butler}, R.~P., {Marcy}, G.~W., {et~al.} 2008, \pasp, 120, 531

\bibitem[{Debes \& Sigurdsson(2002)}]{Debes2002}
Debes, J.~H., \& Sigurdsson, S. 2002, The Astrophysical Journal, 572, 556.
\newblock \url{https://doi.org/10.1086/340291}

\bibitem[{{Debes} {et~al.}(2019){Debes}, {Th{\'e}venot}, {Kuchner},
  {Burgasser}, {Schneider}, {Meisner}, {Gagn{\'e}}, {Faherty}, {Rees}, {Allen},
  {Caselden}, {Cushing}, {Wisniewski}, {Allers}, {Backyard Worlds: Planet 9
  Collaboration}, \& {Disk Detective Collaboration}}]{debes2019}
{Debes}, J.~H., {Th{\'e}venot}, M., {Kuchner}, M.~J., {et~al.} 2019, \apjl,
  872, L25

\bibitem[{{Dones} {et~al.}(2015){Dones}, {Brasser}, {Kaib}, \&
  {Rickman}}]{dones2015}
{Dones}, L., {Brasser}, R., {Kaib}, N., \& {Rickman}, H. 2015, \ssr, 197, 191

\bibitem[{Drell {et~al.}(1965)Drell, Foley, \& Ruderman}]{Drell1965}
Drell, S.~D., Foley, H.~M., \& Ruderman, M.~A. 1965, Journal of Geophysical
  Research (1896-1977), 70, 3131

\bibitem[{{Duncan} {et~al.}(1987){Duncan}, {Quinn}, \& {Tremaine}}]{duncan1987}
{Duncan}, M., {Quinn}, T., \& {Tremaine}, S. 1987, \aj, 94, 1330

\bibitem[{{Duncan} {et~al.}(1988){Duncan}, {Quinn}, \& {Tremaine}}]{duncan1988}
---. 1988, \apjl, 328, L69

\bibitem[{{Everhart}(1967)}]{Everhart67}
{Everhart}, E. 1967, The Astronomical Journal, 72, 1002

\bibitem[{{Fabrycky} {et~al.}(2014){Fabrycky}, {Lissauer}, {Ragozzine}, {Rowe},
  {Steffen}, {Agol}, {Barclay}, {Batalha}, {Borucki}, {Ciardi}, {Ford},
  {Gautier}, {Geary}, {Holman}, {Jenkins}, {Li}, {Morehead}, {Morris},
  {Shporer}, {Smith}, {Still}, \& {Van Cleve}}]{fabrycky2014}
{Fabrycky}, D.~C., {Lissauer}, J.~J., {Ragozzine}, D., {et~al.} 2014, \apj,
  790, 146

\bibitem[{Farihi(2016)}]{Farihi2016}
Farihi, J. 2016, New Astronomy Reviews, 71, 9 .
\newblock
  \url{http://www.sciencedirect.com/science/article/pii/S1387647315300075}

\bibitem[{{Farihi} {et~al.}(2009){Farihi}, {Jura}, \& {Zuckerman}}]{farihi2009}
{Farihi}, J., {Jura}, M., \& {Zuckerman}, B. 2009, \apj, 694, 805

\bibitem[{Farihi {et~al.}(2017)Farihi, von Hippel, \& Pringle}]{Farihi2017}
Farihi, J., von Hippel, T., \& Pringle, J.~E. 2017, Monthly Notices of the
  Royal Astronomical Society: Letters, 471, L145.
\newblock \url{https://doi.org/10.1093/mnrasl/slx122}

\bibitem[{{Fern{\'a}ndez}(1997)}]{fernandez1997}
{Fern{\'a}ndez}, J.~A. 1997, \icarus, 129, 106

\bibitem[{{Fern{\'a}ndez} \& {Ip}(1981)}]{fernandez1981}
{Fern{\'a}ndez}, J.~A., \& {Ip}, W.~H. 1981, \icarus, 47, 470

\bibitem[{{Fern{\'a}ndez} \& {Ip}(1984)}]{fernandez1984}
---. 1984, \icarus, 58, 109

\bibitem[{Ferrario {et~al.}(2015)Ferrario, de~Martino, \&
  G{\"a}nsicke}]{Ferrario2015}
Ferrario, L., de~Martino, D., \& G{\"a}nsicke, B.~T. 2015, Space Science
  Reviews, 191, 111

\bibitem[{{Ferrario} \& {Wickramasinghe}(2005)}]{ferrario2005}
{Ferrario}, L., \& {Wickramasinghe}, D.~T. 2005, \mnras, 356, 615

\bibitem[{Francis(2005)}]{Francis05}
Francis, P.~J. 2005, The Astrophysical Journal, 635, 1348

\bibitem[{{Fraser} {et~al.}(2014){Fraser}, {Brown}, {Morbidelli}, {Parker}, \&
  {Batygin}}]{fraser2014}
{Fraser}, W.~C., {Brown}, M.~E., {Morbidelli}, A., {Parker}, A., \& {Batygin},
  K. 2014, \apj, 782, 100

\bibitem[{{Gomes} {et~al.}(2005){Gomes}, {Levison}, {Tsiganis}, \&
  {Morbidelli}}]{gomes2005}
{Gomes}, R., {Levison}, H.~F., {Tsiganis}, K., \& {Morbidelli}, A. 2005, \nat,
  435, 466

\bibitem[{{Grishin} \& {Veras}(2019)}]{Grishin2019}
{Grishin}, E., \& {Veras}, D. 2019, \mnras, 489, 168

\bibitem[{Hadjidemetriou(1963)}]{Hadjidemetriou1963}
Hadjidemetriou, J.~D. 1963, Icarus, 2, 440 .
\newblock
  \url{http://www.sciencedirect.com/science/article/pii/0019103563900721}

\bibitem[{Hamers \& Portegies~Zwart(2016)}]{Hamers2016}
Hamers, A.~S., \& Portegies~Zwart, S.~F. 2016, Monthly Notices of the Royal
  Astronomical Society: Letters, 462, L84.
\newblock \url{https://doi.org/10.1093/mnrasl/slw134}

\bibitem[{Hanse {et~al.}(2017)Hanse, J\'ilkov\'a, Portegies~Zwart, \&
  Pelupessy}]{Hanse18}
Hanse, J., J\'ilkov\'a, L., Portegies~Zwart, S.~F., \& Pelupessy, F.~I. 2017,
  Monthly Notices of the Royal Astronomical Society, 473, 5432

\bibitem[{{Hayashi} {et~al.}(1985){Hayashi}, {Nakazawa}, \&
  {Nakagawa}}]{Hayashi85}
{Hayashi}, C., {Nakazawa}, K., \& {Nakagawa}, Y. 1985, in Protostars and
  Planets II, ed. D.~C. {Black} \& M.~S. {Matthews}, 1100--1153

\bibitem[{{Heisler} \& {Tremaine}(1986)}]{heisler1986}
{Heisler}, J., \& {Tremaine}, S. 1986, \icarus, 65, 13

\bibitem[{{Hermes} {et~al.}(2017){Hermes}, {G{\"a}nsicke}, {Kawaler}, {Greiss},
  {Tremblay}, {Gentile Fusillo}, {Raddi}, {Fanale}, {Bell}, {Dennihy}, {Fuchs},
  {Dunlap}, {Clemens}, {Montgomery}, {Winget}, {Chote}, {Marsh}, \&
  {Redfield}}]{Hermes2017}
{Hermes}, J.~J., {G{\"a}nsicke}, B.~T., {Kawaler}, S.~D., {et~al.} 2017, \apjs,
  232, 23

\bibitem[{Hestroffer {et~al.}(2019)Hestroffer, S{\'a}nchez, Staron, Bagatin,
  Eggl, Losert, Murdoch, Opsomer, Radjai, Richardson,
  {et~al.}}]{Hestroffer2019}
Hestroffer, D., S{\'a}nchez, P., Staron, L., {et~al.} 2019, The Astronomy and
  Astrophysics Review, 27, 6

\bibitem[{{Hills}(1981)}]{hills1981}
{Hills}, J.~G. 1981, \aj, 86, 1730

\bibitem[{{Hogg} {et~al.}(2021){Hogg}, {Cutter}, \& {Wynn}}]{Hogg2021}
{Hogg}, M.~A., {Cutter}, R., \& {Wynn}, G.~A. 2021, \mnras, 500, 2986

\bibitem[{Hollands {et~al.}(2015)Hollands, Gänsicke, \&
  Koester}]{Hollands2015}
Hollands, M.~A., Gänsicke, B.~T., \& Koester, D. 2015, Monthly Notices of the
  Royal Astronomical Society, 450, 681.
\newblock \url{https://doi.org/10.1093/mnras/stv570}

\bibitem[{Holsapple(2007)}]{Holsapple2007}
Holsapple, K.~A. 2007, Icarus, 187, 500

\bibitem[{Holsapple \& Michel(2008)}]{Holsapple08}
Holsapple, K.~A., \& Michel, P. 2008, {Icarus}, 193, 283

\bibitem[{{Howard} {et~al.}(2012){Howard}, {Marcy}, {Bryson}, {Jenkins},
  {Rowe}, {Batalha}, {Borucki}, {Koch}, {Dunham}, {Gautier}, {Van Cleve},
  {Cochran}, {Latham}, {Lissauer}, {Torres}, {Brown}, {Gilliland}, {Buchhave},
  {Caldwell}, {Christensen-Dalsgaard}, {Ciardi}, {Fressin}, {Haas}, {Howell},
  {Kjeldsen}, {Seager}, {Rogers}, {Sasselov}, {Steffen}, {Basri},
  {Charbonneau}, {Christiansen}, {Clarke}, {Dupree}, {Fabrycky}, {Fischer},
  {Ford}, {Fortney}, {Tarter}, {Girouard}, {Holman}, {Johnson}, {Klaus},
  {Machalek}, {Moorhead}, {Morehead}, {Ragozzine}, {Tenenbaum}, {Twicken},
  {Quinn}, {Isaacson}, {Shporer}, {Lucas}, {Walkowicz}, {Welsh}, {Boss},
  {Devore}, {Gould}, {Smith}, {Morris}, {Prsa}, {Morton}, {Still}, {Thompson},
  {Mullally}, {Endl}, \& {MacQueen}}]{howard2012}
{Howard}, A.~W., {Marcy}, G.~W., {Bryson}, S.~T., {et~al.} 2012, \apjs, 201, 15

\bibitem[{{Hurley} {et~al.}(2000){Hurley}, {Pols}, \& {Tout}}]{hurley2000}
{Hurley}, J.~R., {Pols}, O.~R., \& {Tout}, C.~A. 2000, \mnras, 315, 543

\bibitem[{{Ida} \& {Lin}(2004)}]{ida2004}
{Ida}, S., \& {Lin}, D.~N.~C. 2004, \apj, 604, 388

\bibitem[{{Ida} {et~al.}(2013){Ida}, {Lin}, \& {Nagasawa}}]{ida2013}
{Ida}, S., {Lin}, D.~N.~C., \& {Nagasawa}, M. 2013, \apj, 775, 42

\bibitem[{Isern {et~al.}(2017)Isern, Garc{\'{\i}}a-Berro, Külebi, \&
  Lor{\'{e}}n-Aguilar}]{Isern17}
Isern, J., Garc{\'{\i}}a-Berro, E., Külebi, B., \& Lor{\'{e}}n-Aguilar, P.
  2017, The Astrophysical Journal, 836, L28.
\newblock \url{https://doi.org/10.3847/2041-8213/aa5eae}

\bibitem[{Jura(2003)}]{Jura2003}
Jura, M. 2003, The Astrophysical Journal, 584, L91.
\newblock \url{https://doi.org/10.1086/374036}

\bibitem[{Kaib \& Quinn(2009)}]{Kaib2009}
Kaib, N.~A., \& Quinn, T. 2009, Science, 325, 1234.
\newblock \url{https://science.sciencemag.org/content/325/5945/1234}

\bibitem[{Kawka {et~al.}(2019)Kawka, Vennes, Ferrario, \& Paunzen}]{Kawka2019}
Kawka, A., Vennes, S., Ferrario, L., \& Paunzen, E. 2019, Monthly Notices of
  the Royal Astronomical Society, 482, 5201.
\newblock \url{https://doi.org/10.1093/mnras/sty3048}

\bibitem[{Kenyon \& Bromley(2017)}]{Kenyon2017}
Kenyon, S.~J., \& Bromley, B.~C. 2017, The Astrophysical Journal, 844, 116.
\newblock \url{https://doi.org/10.3847/1538-4357/aa7b85}

\bibitem[{King(1993)}]{King1993}
King, A.~R. 1993, Monthly Notices of the Royal Astronomical Society, 261, 144.
\newblock \url{https://doi.org/10.1093/mnras/261.1.144}

\bibitem[{{Koester} {et~al.}(1998){Koester}, {Dreizler}, {Weidemann}, \&
  {Allard}}]{koester1998}
{Koester}, D., {Dreizler}, S., {Weidemann}, V., \& {Allard}, N.~F. 1998, \aap,
  338, 612

\bibitem[{Koester {et~al.}(2014)Koester, G\"ansicke, \& Farihi}]{Koester2014}
Koester, D., G\"ansicke, B.~T., \& Farihi, J. 2014, \aap, 566, A34.
\newblock \url{https://doi.org/10.1051/0004-6361/201423691}

\bibitem[{Kratter \& Perets(2012)}]{Kratter2012}
Kratter, K.~M., \& Perets, H.~B. 2012, The Astrophysical Journal, 753, 91.
\newblock \url{https://doi.org/10.1088/0004-637x/753/1/91}

\bibitem[{{Levison} {et~al.}(2001){Levison}, {Dones}, \&
  {Duncan}}]{levison2001}
{Levison}, H.~F., {Dones}, L., \& {Duncan}, M.~J. 2001, \aj, 121, 2253

\bibitem[{{Levison} {et~al.}(2006){Levison}, {Duncan}, {Dones}, \&
  {Gladman}}]{levison2006}
{Levison}, H.~F., {Duncan}, M.~J., {Dones}, L., \& {Gladman}, B.~J. 2006,
  \icarus, 184, 619

\bibitem[{{Long} {et~al.}(2018){Long}, {Pinilla}, {Herczeg}, {Harsono},
  {Dipierro}, {Pascucci}, {Hendler}, {Tazzari}, {Ragusa}, {Salyk}, {Edwards},
  {Lodato}, {van de Plas}, {Johnstone}, {Liu}, {Boehler}, {Cabrit}, {Manara},
  {Menard}, {Mulders}, {Nisini}, {Fischer}, {Rigliaco}, {Banzatti}, {Avenhaus},
  \& {Gully-Santiago}}]{long2018}
{Long}, F., {Pinilla}, P., {Herczeg}, G.~J., {et~al.} 2018, \apj, 869, 17

\bibitem[{{Malamud} {et~al.}(2021){Malamud}, {Grishin}, \&
  {Brouwers}}]{Malamud2021}
{Malamud}, U., {Grishin}, E., \& {Brouwers}, M. 2021, \mnras, 501, 3806

\bibitem[{Malamud \& Perets(2020)}]{Malamud2020}
Malamud, U., \& Perets, H.~B. 2020, Monthly Notices of the Royal Astronomical
  Society, 492, 5561.
\newblock \url{https://doi.org/10.1093/mnras/staa142}

\bibitem[{Manser {et~al.}(2020)Manser, Gänsicke, Gentile Fusillo, Ashley,
  Breedt, Hollands, Izquierdo, \& Pelisoli}]{Manser2020}
Manser, C.~J., Gänsicke, B.~T., Gentile Fusillo, N.~P., {et~al.} 2020,
  Monthly Notices of the Royal Astronomical Society, 493, 2127.
\newblock \url{https://doi.org/10.1093/mnras/staa359}

\bibitem[{Manser {et~al.}(2019)Manser, G{\"a}nsicke, Eggl, Hollands, Izquierdo,
  Koester, Landstreet, Lyra, Marsh, Meru, Mustill, Rodr{\'\i}guez-Gil, Toloza,
  Veras, Wilson, Burleigh, Davies, Farihi, Gentile~Fusillo, de~Martino,
  Parsons, Quirrenbach, Raddi, Reffert, Del~Santo, Schreiber, Silvotti, Toonen,
  Villaver, Wyatt, Xu, \& Portegies~Zwart}]{Manser2019}
Manser, C.~J., G{\"a}nsicke, B.~T., Eggl, S., {et~al.} 2019, Science, 364, 66.
\newblock \url{https://science.sciencemag.org/content/364/6435/66}

\bibitem[{{Meech} {et~al.}(2004){Meech}, {Hainaut}, \& {Marsden}}]{meech2004}
{Meech}, K.~J., {Hainaut}, O.~R., \& {Marsden}, B.~G. 2004, \icarus, 170, 463

\bibitem[{{Morbidelli} {et~al.}(2021){Morbidelli}, {Nesvorny}, {Bottke}, \&
  {Marchi}}]{morbidelli2021}
{Morbidelli}, A., {Nesvorny}, D., {Bottke}, W.~F., \& {Marchi}, S. 2021,
  \icarus, 356, 114256

\bibitem[{{Mullally} {et~al.}(2007){Mullally}, {Kilic}, {Reach}, {Kuchner},
  {von Hippel}, {Burrows}, \& {Winget}}]{mullally2007}
{Mullally}, F., {Kilic}, M., {Reach}, W.~T., {et~al.} 2007, \apjs, 171, 206

\bibitem[{{Nagasawa} {et~al.}(2005){Nagasawa}, {Lin}, \&
  {Thommes}}]{nagasawa2005}
{Nagasawa}, M., {Lin}, D.~N.~C., \& {Thommes}, E. 2005, \apj, 635, 578

\bibitem[{{Neslu{\v{s}}an}(2007)}]{Neslusan07}
{Neslu{\v{s}}an}, L. 2007, Astronomy \& Astrophysics, 461, 741

\bibitem[{Neubauer(1980)}]{Neubauer1980}
Neubauer, F. 1980, Journal of Geophysical Research: Space Physics, 85, 1171.
\newblock \url{https://doi.org/10.1029/JA085iA03p01171}

\bibitem[{{O'Connor} \& {Lai}(2020)}]{OConnor2020}
{O'Connor}, C.~E., \& {Lai}, D. 2020, \mnras, 498, 4005.
\newblock \url{https://ui.adsabs.harvard.edu/abs/2020MNRAS.498.4005O}

\bibitem[{{Oort}(1950)}]{oort1950}
{Oort}, J.~H. 1950, \bain, 11, 91

\bibitem[{{Petigura} {et~al.}(2013){Petigura}, {Marcy}, \&
  {Howard}}]{petigura2013}
{Petigura}, E.~A., {Marcy}, G.~W., \& {Howard}, A.~W. 2013, \apj, 770, 69

\bibitem[{Pohl \& Britt(2020)}]{Pohl2020}
Pohl, L., \& Britt, D.~T. 2020, Meteoritics \& Planetary Science, 55, 962.
\newblock \url{https://doi.org/10.1111/maps.13449}

\bibitem[{Portegies~Zwart {et~al.}(2018)Portegies~Zwart, Torres, Pelupessy,
  Bédorf, \& Cai}]{Portegies18}
Portegies~Zwart, S., Torres, S., Pelupessy, I., Bédorf, J., \& Cai, M.~X.
  2018, Monthly Notices of the Royal Astronomical Society: Letters, 479, L17

\bibitem[{{Punzo} {et~al.}(2014){Punzo}, {Capuzzo-Dolcetta}, \& {Portegies
  Zwart}}]{punzo2014}
{Punzo}, D., {Capuzzo-Dolcetta}, R., \& {Portegies Zwart}, S. 2014, \mnras,
  444, 2808

\bibitem[{Rafikov(2011)}]{Rafikov2011}
Rafikov, R.~R. 2011, The Astrophysical Journal, 732, L3.
\newblock \url{https://doi.org/10.1088/2041-8205/732/1/l3}

\bibitem[{Rafikov(2018)}]{Rafikov18}
---. 2018, The Astrophysical Journal, 861, 35

\bibitem[{Rafikov \& Garmilla(2012)}]{Rafikov2012}
Rafikov, R.~R., \& Garmilla, J.~A. 2012, The Astrophysical Journal, 760, 123.
\newblock \url{https://doi.org/10.1088/0004-637x/760/2/123}

\bibitem[{{Rein} \& {Liu}(2012)}]{Rein2012}
{Rein}, H., \& {Liu}, S.~F. 2012, \aap, 537, A128

\bibitem[{{Rein} \& {Spiegel}(2015)}]{Rein2015}
{Rein}, H., \& {Spiegel}, D.~S. 2015, \mnras, 446, 1424

\bibitem[{Richardson {et~al.}(2002)Richardson, Leinhardt, Melosh, \&
  Michel}]{Richardson2002}
Richardson, D.~C., Leinhardt, Z.~M., Melosh, H.~J., \& Michel, P. 2002, in
  Asteroids III, ed. W.~F. Bottke, A.~Cellino, P.~Paolicchi, \& P.~R. Binzel
  (Tucson: Univ. of Arizona), 501--515

\bibitem[{Scheeres {et~al.}(2010)Scheeres, Hartzell, Sánchez, \&
  Swift}]{Scheeres2010}
Scheeres, D., Hartzell, C., Sánchez, P., \& Swift, M. 2010, Icarus, 210, 968

\bibitem[{Schr{\"o}der \& Connon~Smith(2008)}]{Schroder2008}
Schr{\"o}der, K.-P., \& Connon~Smith, R. 2008, Monthly Notices of the Royal
  Astronomical Society, 386, 155.
\newblock \url{https://doi.org/10.1111/j.1365-2966.2008.13022.x}

\bibitem[{{Spurzem} {et~al.}(2009){Spurzem}, {Giersz}, {Heggie}, \&
  {Lin}}]{spurzem2009}
{Spurzem}, R., {Giersz}, M., {Heggie}, D.~C., \& {Lin}, D.~N.~C. 2009, \apj,
  697, 458

\bibitem[{Sridhar \& Tremaine(1992)}]{Sridhar1992}
Sridhar, S., \& Tremaine, S. 1992, Icarus, 95, 86 .
\newblock
  \url{http://www.sciencedirect.com/science/article/pii/001910359290193B}

\bibitem[{{Tamayo} {et~al.}(2020){Tamayo}, {Rein}, {Shi}, \&
  {Hernandez}}]{Tamayo2020}
{Tamayo}, D., {Rein}, H., {Shi}, P., \& {Hernandez}, D.~M. 2020, \mnras, 491,
  2885

\bibitem[{Tremblay {et~al.}(2016)Tremblay, Cummings, Kalirai, Gänsicke,
  Gentile-Fusillo, \& Raddi}]{Tremblay2016}
Tremblay, P.-E., Cummings, J., Kalirai, J.~S., {et~al.} 2016, Monthly Notices
  of the Royal Astronomical Society, 461, 2100.
\newblock \url{https://doi.org/10.1093/mnras/stw1447}

\bibitem[{{Tsiganis} {et~al.}(2005){Tsiganis}, {Gomes}, {Morbidelli}, \&
  {Levison}}]{tsiganis2005}
{Tsiganis}, K., {Gomes}, R., {Morbidelli}, A., \& {Levison}, H.~F. 2005, \nat,
  435, 459

\bibitem[{Vanderbosch {et~al.}(2020)Vanderbosch, Hermes, Dennihy, Dunlap,
  Izquierdo, Tremblay, Cho, Gänsicke, Toloza, Bell, Montgomery, \&
  Winget}]{Vanderbosch2020}
Vanderbosch, Z., Hermes, J.~J., Dennihy, E., {et~al.} 2020, The Astrophysical
  Journal, 897, 171.
\newblock \url{https://doi.org/10.3847/1538-4357/ab9649}

\bibitem[{Vanderburg {et~al.}(2015)Vanderburg, Johnson, Rappaport, Bieryla,
  Irwin, Lewis, Kipping, Brown, Dufour, Ciardi, {et~al.}}]{Vanderburg2015}
Vanderburg, A., Johnson, J.~A., Rappaport, S., {et~al.} 2015, Nature, 526, 546

\bibitem[{Veras(2016)}]{Veras2016a}
Veras, D. 2016, Royal Society Open Science, 3, 150571

\bibitem[{Veras {et~al.}(2015{\natexlab{a}})Veras, Eggl, \&
  Gänsicke}]{Veras2015b}
Veras, D., Eggl, S., \& Gänsicke, B.~T. 2015{\natexlab{a}}, Monthly Notices of
  the Royal Astronomical Society, 452, 1945.
\newblock \url{https://doi.org/10.1093/mnras/stv1417}

\bibitem[{Veras \& Heng(2020)}]{Veras2020}
Veras, D., \& Heng, K. 2020, Monthly Notices of the Royal Astronomical Society,
  496, 2292.
\newblock \url{https://doi.org/10.1093/mnras/staa1632}

\bibitem[{Veras {et~al.}(2019)Veras, Higuchi, \& Ida}]{Veras2019a}
Veras, D., Higuchi, A., \& Ida, S. 2019, Monthly Notices of the Royal
  Astronomical Society, 485, 708.
\newblock \url{https://doi.org/10.1093/mnras/stz421}

\bibitem[{Veras {et~al.}(2014)Veras, Leinhardt, Bonsor, \&
  G{\"a}nsicke}]{Veras2014}
Veras, D., Leinhardt, Z.~M., Bonsor, A., \& G{\"a}nsicke, B.~T. 2014, Monthly
  Notices of the Royal Astronomical Society, 445, 2244.
\newblock \url{https://doi.org/10.1093/mnras/stu1871}

\bibitem[{Veras {et~al.}(2015{\natexlab{b}})Veras, Leinhardt, Eggl, \&
  G{\"a}nsicke}]{Veras2015a}
Veras, D., Leinhardt, Z.~M., Eggl, S., \& G{\"a}nsicke, B.~T.
  2015{\natexlab{b}}, Monthly Notices of the Royal Astronomical Society, 451,
  3453.
\newblock \url{https://doi.org/10.1093/mnras/stv1195}

\bibitem[{Veras \& Wolszczan(2019)}]{Veras2019b}
Veras, D., \& Wolszczan, A. 2019, Monthly Notices of the Royal Astronomical
  Society, 488, 153.
\newblock \url{https://doi.org/10.1093/mnras/stz1721}

\bibitem[{{Veras} {et~al.}(2011){Veras}, {Wyatt}, {Mustill}, {Bonsor}, \&
  {Eldridge}}]{veras2011}
{Veras}, D., {Wyatt}, M.~C., {Mustill}, A.~J., {Bonsor}, A., \& {Eldridge},
  J.~J. 2011, \mnras, 417, 2104

\bibitem[{{Veras} {et~al.}(2020){Veras}, {Reichert}, {Flammini Dotti}, {Cai},
  {Mustill}, {Shannon}, {McDonald}, {Portegies Zwart}, {Kouwenhoven}, \&
  {Spurzem}}]{Veras2020a}
{Veras}, D., {Reichert}, K., {Flammini Dotti}, F., {et~al.} 2020, \mnras, 493,
  5062

\bibitem[{von Hippel {et~al.}(2007)von Hippel, Kuchner, Kilic, Mullally, \&
  Reach}]{von_Hippel2007}
von Hippel, T., Kuchner, M.~J., Kilic, M., Mullally, F., \& Reach, W.~T. 2007,
  The Astrophysical Journal, 662, 544.
\newblock \url{https://doi.org/10.1086/518108}

\bibitem[{Walsh {et~al.}(2019)Walsh, Jawin, Ballouz, Barnouin, Bierhaus,
  Connolly, Molaro, McCoy, Delbo, Hartzell, \& Lauretta}]{Walsh2019}
Walsh, K.~J., Jawin, E.~R., Ballouz, R.-L., {et~al.} 2019, Nature Geoscience,
  12, 242

\bibitem[{{Weissman}(1983)}]{weissman1983}
{Weissman}, P.~R. 1983, \aap, 118, 90

\bibitem[{{Weissman} \& {Levison}(1997)}]{Weissman97}
{Weissman}, P.~R., \& {Levison}, H.~F. 1997, The Astrophysical Journal, 488,
  L133

\bibitem[{{Wiegert} \& {Tremaine}(1999)}]{wiegert1999}
{Wiegert}, P., \& {Tremaine}, S. 1999, \icarus, 137, 84

\bibitem[{Xu {et~al.}(2017)Xu, Rappaport, van Lieshout, Vanderburg, Gary,
  Hallakoun, Ivanov, Wyatt, DeVore, Bayliss, Bento, Bieryla, Cameron, Cann,
  Croll, Collins, Dalba, Debes, Doyle, Dufour, Ely, Espinoza, Joner, Jura,
  Kaye, McClain, Muirhead, Palle, Panka, Provencal, Randall, Rodriguez,
  Scarborough, Sefako, Shporer, Strickland, Zhou, \& Zuckerman}]{Xu2017}
Xu, S., Rappaport, S., van Lieshout, R., {et~al.} 2017, Monthly Notices of the
  Royal Astronomical Society, 474, 4795.
\newblock \url{https://doi.org/10.1093/mnras/stx3023}

\bibitem[{Zhang \& Lin(2020)}]{ZhangLin2020}
Zhang, Y., \& Lin, D. N.~C. 2020, Nature astronomy, 4, 852

\bibitem[{Zhang \& Michel(2020)}]{ZhangMichel2020}
Zhang, Y., \& Michel, P. 2020, Astronomy \& Astrophysics, 640, A102

\bibitem[{Zhang {et~al.}(2018)Zhang, Richardson, Barnouin, Michel, Schwartz, \&
  Ballouz}]{Zhang2018}
Zhang, Y., Richardson, D.~C., Barnouin, O.~S., {et~al.} 2018, The Astrophysical
  Journal, 857, 15

\bibitem[{Zhang {et~al.}(2017)Zhang, Richardson, Barnouin, Maurel, Michel,
  Schwartz, Ballouz, Benner, Naidu, \& Li}]{Zhang2017}
---. 2017, Icarus, 294, 98

\bibitem[{{Zheng} {et~al.}(2017{\natexlab{a}}){Zheng}, {Lin}, \&
  {Kouwenhoven}}]{zheng2017a}
{Zheng}, X., {Lin}, D. N.~C., \& {Kouwenhoven}, M.~B.~N. 2017{\natexlab{a}},
  \apj, 836, 207

\bibitem[{{Zheng} {et~al.}(2017{\natexlab{b}}){Zheng}, {Lin}, {Kouwenhoven},
  {Mao}, \& {Zhang}}]{zheng2017}
{Zheng}, X., {Lin}, D. N.~C., {Kouwenhoven}, M.~B.~N., {Mao}, S., \& {Zhang},
  X. 2017{\natexlab{b}}, \apj, 849, 98

\bibitem[{Zhou \& Lin(2007)}]{Zhoulin07}
Zhou, J.-L., \& Lin, D. N.~C. 2007, The Astrophysical Journal, 666, 447

\bibitem[{Zuckerman \& Becklin(1987)}]{Zuckerman1987}
Zuckerman, B., \& Becklin, E. 1987, Nature, 330, 138

\bibitem[{Zuckerman {et~al.}(2003)Zuckerman, Koester, Reid, \&
  Hunsch}]{Zuckerman2003}
Zuckerman, B., Koester, D., Reid, I.~N., \& Hunsch, M. 2003, The Astrophysical
  Journal, 596, 477.
\newblock \url{https://doi.org/10.1086/377492}

\end{thebibliography}
\end{CJK*} 

\end{document}